\documentclass[showpacs,groupedaddress,twocolumn,aps,pra]{revtex4-1}
\usepackage{amsmath}
\usepackage{amssymb}
\usepackage{graphicx}    
\usepackage[normalem]{ulem}

\renewcommand{\v}[1]{\ensuremath{\mathbf{#1}}} 
\newcommand{\gv}[1]{\ensuremath{\boldsymbol{ #1}} } 
\newcommand{\expv}[1]{\left\langle #1 \right\rangle}	
\newcommand{\arctanh}{\operatorname{arctanh}}
\newcommand{\grad}[1]{\gv{\nabla} #1} 

\begin{document}         
\title{Decay of a superfluid current of ultra-cold atoms in a toroidal trap}
\author{Amy~C.~Mathey$^{1}$, Charles~W.~Clark$^{2}$, L.~Mathey$^{1,3}$}
\affiliation{
$^1$Zentrum f\"ur Optische Quantentechnologien and Institut f\"ur Laserphysik, Universit\"at Hamburg, 22761 Hamburg, Germany\\
$^2$Joint Quantum Institute, National Institute of Standards and Technology \& University of Maryland, Gaithersburg, MD 20899\\
$^3$The Hamburg Centre for Ultrafast Imaging, Luruper Chaussee 149, Hamburg 22761, Germany}

\date{\today}

\begin{abstract}
Using a numerical implementation of the truncated Wigner approximation, we simulate the experiment reported by Ramanathan {\it et al.} in Phys.~Rev.~Lett.~\textbf{106}, 130401 (2011), in which a Bose-Einstein condensate is created in a toroidal trap and set into rotation via a phase imprinting technique. A potential barrier is then placed in the trap to study the decay of the superflow. We find that the current decays via thermally activated phase slips, which can also be visualized as vortices crossing the barrier region in the radial direction.  Adopting the notion of  critical velocity used in the experiment, we determine it to be lower than the local speed of sound at the barrier, in contradiction to the predictions of the zero-temperature Gross-Pitaevskii equation. We map out the superfluid decay rate and critical velocity as a function of temperature and observe a strong dependence.  Thermal fluctuations offer a partial explanation of the experimentally observed reduction of the critical velocity from the phonon velocity.

\end{abstract}

\maketitle

\section{Introduction}\label{sec:intro}

Superfluidity is a compelling and counter-intuitive phenomenon that has intrigued scientists for decades. The interplay of quantum motion of particles, quantum statistics and interactions gives rise to dissipationless flow, the defining property of superfluidity. This flow, however, will only be sustained within a certain parameter regime. If the system experiences a sufficiently large perturbation, its dissipationless nature will break down. To understand this breakdown in fact constitutes understanding superfluidity itself, as it entails understanding why excitations are suppressed in the superfluid regime and what constitutes a sufficiently large perturbation that will destroy superfluidity.

Experiments in superfluid helium, Refs.~\cite{rayfield_roton_1966,*kukich_decay_1968, *anderson_considerations_1966, *varoquaux_phase_1994, *packard_phase_1994,langer_intrinsic_1967-1}, seem to suggest that one such perturbation is an impurity or container wall that moves relative to the superfluid with sufficiently large speed that it leads to a breakdown of superfluidity. This indicates the possibility of a critical velocity, above which dissipation develops and the superfluid current decays. In a seminal study, Landau related the critical velocity to the elementary excitations \cite{landau_theory_1941} of the system. An excitation of energy $\epsilon(p)$ with momentum $p$ can only be created above the velocity $v_c= \min (\epsilon(p)/|p|)$, while fulfilling both energy and momentum conservation. For a system with an excitation spectrum which has a roton minimum, such as helium, the excitation of rotons determines the critical velocity of superfluid helium. For a weakly interacting system with a Bogoliubov excitation spectrum, the low-energy excitations are phonons with energy $\epsilon(k)=\hbar c |k|$, where $k$ is the wave number, and the above expression is equal to the speed of sound $c$.  
 Feynman considered yet another type of excitation, in the situation where a superfluid flows out of a channel into a reservoir and suggested that the relevant excitations were vortex-anti-vortex pairs \cite{feynman_application_1955}.  Using energetic considerations, he estimated the critical velocity to be $v_c= [ \hbar/(md)] \log(d/a) $, where  $d$ is the channel diameter, $m$ is the atomic mass, and $a$ is the vortex core diameter.  However, many questions about the phenomenon of superfluidity remain unanswered, in particular regarding the effects of the dimensionality of the system, temperature and the boundaries.

With the advances in ultra-cold atom technology these questions can now be addressed in a widely tunable environment, in the flow of Bose-Einstein condensates (BECs), see e.g.~Refs.~\cite{raman_evidence_1999, *onofrio_observation_2000, chikkatur_suppression_2000, burger_superfluid_2001, ramanathan_superflow_2011,desbuquois_superfluid_2012}.  The critical velocity that was found in \cite{raman_evidence_1999,*onofrio_observation_2000} was much smaller than the sound velocity,
 while the ones that were found in \cite{chikkatur_suppression_2000, *burger_superfluid_2001} were comparable to it. Theoretical studies were reported in Refs. \cite{frisch_transition_1992,*jackson_vortex_1998, *jackson_dissipation_2000, *polkovnikov_decay_2004, watanabe_critical_2009, piazza_vortex-induced_2009,piazza_instability_2011, dubessy_critical_2012}. In \cite{watanabe_critical_2009} it was found that for a rectangular barrier the critical velocity is the local sound velocity at the barrier, within a Gross-Pitaevskii equation (GPE) approach in one dimension (1D). In \cite{dubessy_critical_2012} the instability of the flow due to surface modes was explored.

In the experiment performed at NIST \cite{ramanathan_superflow_2011}, a critical velocity less than the local sound speed was found when a barrier was raised into the superfluid flow in a toroidal trap.  
Toroidal BECs, which have been proposed and investigated using a  variety of methods \cite{gupta_bose-einstein_2005,*arnold_large_2006,*morizot_ring_2006, *olson_cold-atom_2007, *sherlock_time-averaged_2011,ryu_observation_2007, moulder_quantized_2012}, have recently been used to generate persistent currents and study their subsequent decay \cite{ryu_observation_2007,ramanathan_superflow_2011, moulder_quantized_2012}.  Potential applications of toroidal BECs include high precision interferometry \cite{gustavson_precision_1997,*lenef_rotation_1997,*wang_atom_2009,*halkyard_rotational_2010} and analogs of SQUIDS in atomtronic circuits \cite{seaman_atomtronics:_2007}.


In this paper, we study the superfluid properties of BECs in toroidal traps using a numerical implementation of the Truncated Wigner approximation (TWA), Refs.~\cite{walls_quantum_2008,polkovnikov_evolution_2003, blakie_dynamics_2008, *polkovnikov_phase_2010}.
 This formalism includes the next order of thermal and quantum fluctuations beyond the GPE-approximation. We simulate the experiment in Ref.~\cite{ramanathan_superflow_2011}, and find that a GPE description is inconsistent with the experimental results. The TWA approach, however, suggests that thermal fluctuations are of visible importance, and further, it allows for the identification of  the decay mechanism, which are phase slips resulting from vortices crossing the barrier region, as we discuss in this paper. We demonstrate the strong temperature dependence of several key observables, which highlights the importance of thermal fluctuations. The comparison to the experimental results suggests that the findings of Ref.~\cite{ramanathan_superflow_2011} constitute `post-GPE' dynamics, in the sense that the inclusion of fluctuations is vital for its understanding.

This paper is organized as follows: In Sect.~\ref{sec:methods} we describe how the system is modeled in our formalism; in Sect.~\ref{sec:crit-vel} we discuss the properties of the superfluid decay that we find; in Sect.~\ref{sec:phase_slip} we illustrate the properties of the phase slip mechanism; and in Sect.~\ref{sec:expt} we compare our results directly to the experimental measurements. In Sect.~\ref{sec:temp} we discuss the temperature dependence of the decay, in Sect.~\ref{sec:conclusion} we conclude. In Appendix \ref{appendix:temperature}, we report our numerical method of determining the temperature of the ensemble. In Appendix \ref{appendix:chem_potential}, we discuss different estimators of the chemical potential and in Appendix \ref{appendix:sound_speed} the dependence of the local speed of sound on dimensionality.

\section{Modeling the system}\label{sec:methods}
%
The semi-classical TWA method was developed in the field of quantum optics \cite{walls_quantum_2008} and later formulated within a path-integral formalism \cite{polkovnikov_evolution_2003}. In this method, an ensemble of initial conditions is generated from the Wigner distribution of the initial state and then propagated according to the classical equations of motion. Observables are calculated in each realization and then averaged over.  This method captures the next order of quantum and thermal fluctuations beyond GPE. 
 Other TWA  studies on ultracold atom systems have been reported on dipolar oscillations \cite{polkovnikov_effect_2004}, non-adiabatic loading of a BEC into an optical lattice \cite{isella_nonadiabatic_2005, *isella_quantum_2006}, dynamics of two-dimensional superfluid bi-layers \cite{mathey_supercritical_2009,*mathey_light_2010,mathey_dynamic_2011}, dynamical instabilities of a BEC in a one-dimensional lattice \cite{ferris_dynamical_2008} and  dynamics of spinor condensates \cite{sau_theory_2009, *barnett_prethermalization_2011}. 
 
 To carry out the numerical simulations, it is convenient to discretize real space and represent the continuous Hamiltonian by the discrete Bose-Hubbard   Hamiltonian \cite{jaksch_cold_1998} on a 3D square lattice of dimensions $N_x \times N_y \times N_z$:
\begin{eqnarray}
 \hat{H} &=& -J\sum_{\langle i j \rangle} \left(  \hat{\psi}_i^\dagger \hat{\psi}_j + \hat{\psi}_j^\dagger \hat{\psi}_i \right) + \frac{U}{2}\sum_i \hat{n}_i(\hat{n}_i-1)\nonumber\\
 	&& + \sum_i V_i(t)\hat{n}_i,\label{eqn:Hamiltonian} 
\end{eqnarray}
where $J$ is the hopping parameter, $U$ is the on-site energy, $\hat{\psi}_j^\dagger (\hat{\psi}_j)$ are the bosonic creation (annihilation) operators at site $j$, and  
  $\langle i j\rangle$ indicates nearest-neighbor bonds. For a lattice discretization length $l$, the Bose-Hubbard parameters are related to the continuum parameters, cp. Ref.~\cite{mora_extension_2003}, by $ J=\hbar^2/(2ml^2)$ and $U=gl^{-3}$, where $m$ is the atom mass, and  $g=4\pi a_s\hbar^2/m$; $a_s$ is the s-wave scattering length.
 The real space location ${\bf r} = (x,y,z)$  is related to the lattice location ${\bf i} = (i_x, i_y, i_z)$ through ${\bf r} = l {\bf i}$.    $L_x=lN_x$, $L_y = lN_y$, $L_z=lN_z$ are the dimensions of the discretized space used in the simulations. 
 To represent approximately the toroidal geometry we use periodic boundary conditions along the $x$-direction. The $y$ direction represents the radial direction and the $z$ direction is the direction perpendicular to the plane of the torus, which corresponds to the vertical direction in the experiment. The origin is located at $(L_x/2, L_y/2, L_z/2)$.  
The cylindrical geometry in our simulations is a good approximation to the toroidal geometry of the experiment because the superfluid decay is governed by the phase slip dynamics at the barrier where the transverse extent of the condensate is small.  Indeed in the GPE simulations, the system is nearly single channel at the critical barrier while in the TWA simulations it is few channel. 
Furthermore, in the TWA simulations, the thermal fluctuations of the velocity are significantly higher than the velocity difference expected for a toriodal geometry, if a fully condensed Thomas-Fermi profile is assumed.
 
  The time-dependent external potential $V_i(t) =  V_{tr, i}(t) + V_{b, i}(t)$ consists of the harmonic trap,  $V_{tr, i}(t) = \alpha(t) \left(\omega_y^2y^2 + \omega_z^2z^2\right)/4Jl^2$, with trapping frequencies $\omega_y$ and $\omega_z$ and a Gaussian barrier potential, $V_{b, i}(t)= \eta(t)V_{b0} \exp\left[-(x - x_b)^2/2l_b^2\right]$. 
  $V_{b0}$ is the strength of the barrier, $x_b$ is its location, and $l_b$ its width. 
  The time-dependent coefficients, $\alpha(t), \eta(t)$ are varied between $0$ and $1$: The trapping potential is ramped up adiabatically to create the initial state, as described below, and the barrier potential is ramped up similarly to the experimental procedure. As discussed in Ref.~\cite{mora_extension_2003}, the discrete model approximates the continuum system when the healing length $\xi\equiv\sqrt{\hbar^2/mgn_{0, i}}$ at some location $i$ and the thermal de Broglie wavelength, $\lambda = \sqrt{2\pi\hbar^2/mk_B T}$ are comparable to or larger than the lattice spacing $l$, where $n_{0, i} = \langle \hat{n}_i\rangle$ is the density, $k_B$ is the Boltzmann constant, and $T$ the temperature. 

Within the TWA approach, the operators $\hat{\psi}_i (t)$ are replaced by classical fields $\psi_i(t)$, which propagate according to
 the equations of motion derived from Eq.~\ref{eqn:Hamiltonian}. We initialize the fields $\psi_i(t=0)$ according to the Wigner distribution of a homogeneous  (i.e. $V_i(t=0) = 0$),  weakly interacting Bose gas, within the Bogoliubov approximation, as in Ref.~\cite{mathey_dynamic_2011}. Using the Bogoliubov transformation in the phase-density representation \cite{mora_extension_2003}, the Hamiltonian (\ref{eqn:Hamiltonian}) is mapped to $\hat{H}^\prime = \sum_\nu \epsilon_\nu \hat{c}^\dagger_\nu \hat{c}_\nu$, where $\epsilon_\nu$ is the energy and $\hat{c}_{\nu}^\dagger (\hat{c}_{\nu})$ are the creation (annihilation) operators for the Bogoliubov modes. Terms beyond quadratic order  in the fluctuations are ignored. The Wigner distribution for a thermal ensemble of harmonic oscillators  at temperature $T_0$ is a product of Gaussians, $W \sim \prod_{\nu} e^{-x_\nu^2/2\sigma_{x,\nu}^2 - p_\nu^2/2\sigma_{p,\nu}^2 } $ with variance $\sigma_{x,\nu}^2 = 1/\left[2\epsilon_{\nu}\tanh(\epsilon_{\nu}/2T_{0}) \right]$ and $\sigma_{p,\nu}^2 = \epsilon_\nu /\left[2\tanh(\epsilon_{\nu}/2T_{0}) \right]$ for the position and momentum, which are mapped to the Bogoliubov modes using $ c_\nu =  \left(i/\sqrt{2\epsilon_\nu} \right) p_\nu + \left( \sqrt{\epsilon_\nu/2}\right) x_\nu$.  
 %
\begin{figure}
\centerline{\includegraphics[scale=0.55]{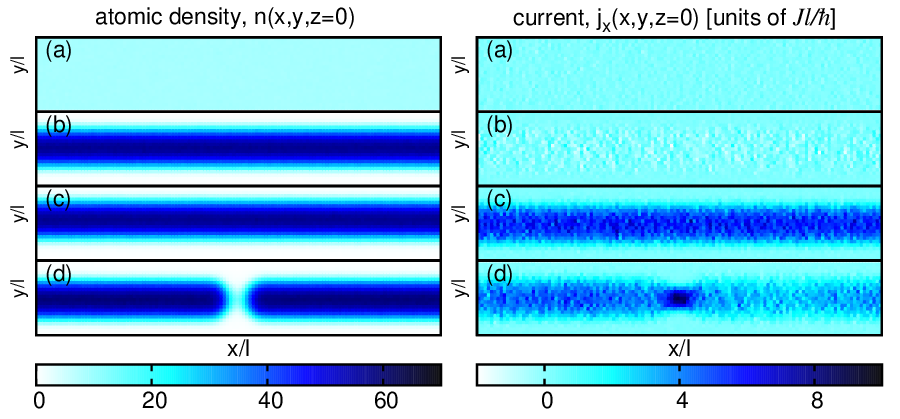}}
\caption{\label{fig:n_jx_sketch}(Color online). Density and current in $x$-direction in the $x-y$ plane at $z=0$. Panel (a) shows the homogeneous ensemble after initialization at $t_n=100\hbar/J$ and $T_0 = 1 J = 10.8 nK$; (b) shows the ensemble after ramping up the trapping potential at $t_n=27 080 \hbar/J$; (c) shows the system after phase imprinting at $t_n=27120 \hbar/J$ and (d) after ramping up a barrier, centered at $x_b=0$, at $t_n=28 000 \hbar/J$. The total lattice size is given by $N_x=126$,  $N_y=21$ and $N_z=7$. The total atom number is $N= 150028$. The barrier strength is $V_{b0} = 2.65 J = 0.67 \mu_0$. The ensemble is initialized with $T_0=1 J$; after turning on the trap the temperature is $T=(5.48 \pm 0.11) J = (59.2 \pm 1.2)$ nK.  The data shown here is averaged over 512 realizations.}
\end{figure}
 After the initialization, the harmonic trapping potential is slowly ramped-up to generate the ensemble in the trap. 
We next measure the temperature of the trapped ensemble, as described in Appendix \ref{appendix:temperature}. Both the approximation of the Wigner distribution and the ramp-up of the external potential lead to heating of the ensemble to a temperature above $T_0$.  
After this the experimental procedure of Ref.~\cite{ramanathan_superflow_2011} is simulated. We imprint a phase winding  $\psi(x)\rightarrow e^{-i\phi(x)}\psi(x)$, with $\phi(x)=2\pi x/L_x$, then ramp up the barrier, hold it constant for approximately $2$s, and ramp it down. 
 	
In Fig.~\ref{fig:n_jx_sketch} we show an example of this process to demonstrate the preparation sequence. In this particular example, we choose $N_x = 126$, $N_y=21$ and $N_z=7$; $n_0 =8.1$ is the expectation value of the density of the initial homogeneous ensemble; $T_0/J = 1$ is the temperature of the initial state. 
We set $U=0.07J$, which translates into a lattice spacing of $l = 0.987$ $\mu$m, a time step $\Delta t=\hbar/J= 0.706$ ms and energy scale $J =10.8$ nK for sodium atoms. 
In Fig.~\ref{fig:n_jx_sketch}(a) we show the homogeneous density in the $x\text{-}y$ plane at $z=0$ and the azimuthal current, defined as
$j_x({\bf r}) = -i J l \hbar^{-1}\left[\psi_{\bf r}^\dagger \psi_{\bf{r}+l\hat{x}} - \psi_{\bf{r}+l\hat{x}}^\dagger\psi_{\bf{r}} \right]$,  at $z=0$. The current $j_x({\bf r})$ has a zero expectation value in the initial state.
Both quantities are averaged over 512 realizations. 
Next, we slowly ramp up the harmonic trap, $V_{tr, i}(t_n)$, according to $\alpha(t_n) = \left\lbrace 1- \tanh[(t_n-t_{0tr})/\tau_{tr}]\right\rbrace/2$, where $\tau_{tr}= 3200 \hbar/J= 2.26$ s  and $t_{0tr}=8100 \hbar/J= 5.72$ s. 
The trapping frequencies in the $y-$ and $z-$ directions are $\hbar\omega_y= 0.5 J = 2\pi \times 113$ Hz, and $\hbar\omega_z=2.5J = 2\pi \times 563$ Hz.  In Fig.~\ref{fig:n_jx_sketch}(b) we show $n(x,y,0)$ and $j_x(x,y,0)$ at time $t_n = 27080 \hbar/J$ after the trap is fully ramped on and the system has been allowed to equilibrate. 
 The initialization in this example is completed at $27000 \hbar/J$.  We introduce the time variable $t=t_n-27000\hbar/J$, which corresponds to the time of the experiment, while $t_n$ corresponds to the numerical time, including the initialization process.
 After the trap has been ramped up, the density is inhomogeneous and has a maximum at the center in $y$-direction. Although the current has a zero expectation value, some fluctuations are visible due to the finite temperature that has been introduced by the initialization process and the trap ramp-up. Using the temperature measurement of Appendix \ref{appendix:temperature}, we determine the temperature to be $T = (5.48 \pm 0.11) J$ for the example in Fig.~\ref{fig:n_jx_sketch}. 

Next, we imprint a $2\pi$ phase winding at time $t_n=27100 \hbar/J$. The density and current just after phase imprinting, at time $t_n = 27120 \hbar/J$ is depicted in Fig.~\ref{fig:n_jx_sketch}(c). The density profile is unchanged, but the current now has a finite expectation value, as the atoms circulate to the right, and displays some thermal fluctuations. 
 A  barrier potential is ramped up at $300 \hbar/J \approx 0.2$ s after phase imprinting.     
 The barrier is centered at $x_b=0$, has a $1/e^2$ width of $2l_b = 6l$, and is ramped up linearly as $ \eta(t_n) =(t_n-t_{0b})/\Delta t_b$ over a time $\Delta t_b = 145 \hbar/J \approx 0.1$s starting at $t_{0b}= 27400\hbar/J$.  For the example in Fig.~\ref{fig:n_jx_sketch}, the barrier strength is $V_{b0} = 2.65 J = 0.67 \mu_0$. The barrier is held at its maximum height for  $2850 \hbar/J \approx 2$ s and ramped back down linearly over $\Delta t_b$.  These time scales are based on the experimental procedure. The density depletion at the barrier is apparent in Fig.~\ref{fig:n_jx_sketch}(d).   Simultaneously, due to the constriction at the barrier, the current at $z=0$ increases at the barrier, while the total current, i.e. the current   integrated over $y$ and $z$, is unchanged. 
  The total ``experiment" time following the initialization is $3600 \hbar/J \approx 2.5$ s. All the numerical results presented in this paper use the same lattice discretization and times described here.

In the following, we discuss numerous simulations, in which the parameters of the system are varied. The total number of atoms ranges from 50,000 to 180,000. The number of lattice sites in $y$- and $z$-directions are chosen to be larger than the Thomas-Fermi radii of the condensate in these directions, and therefore vary with the total number of atoms.  $L_y$ ranges from $17-21l$ and $L_z$ ranges from $5-7l$.  In the elongated direction, the length is $L_x = 126 l = 124.4$ $\mu$m. For a ring with circumference $L_x$, the radius is $R= 19.8$ $\mu$m. 

\section{Decay of Current \& Critical Velocity}\label{sec:crit-vel}
\begin{figure}
\centerline{\includegraphics[scale=0.55]{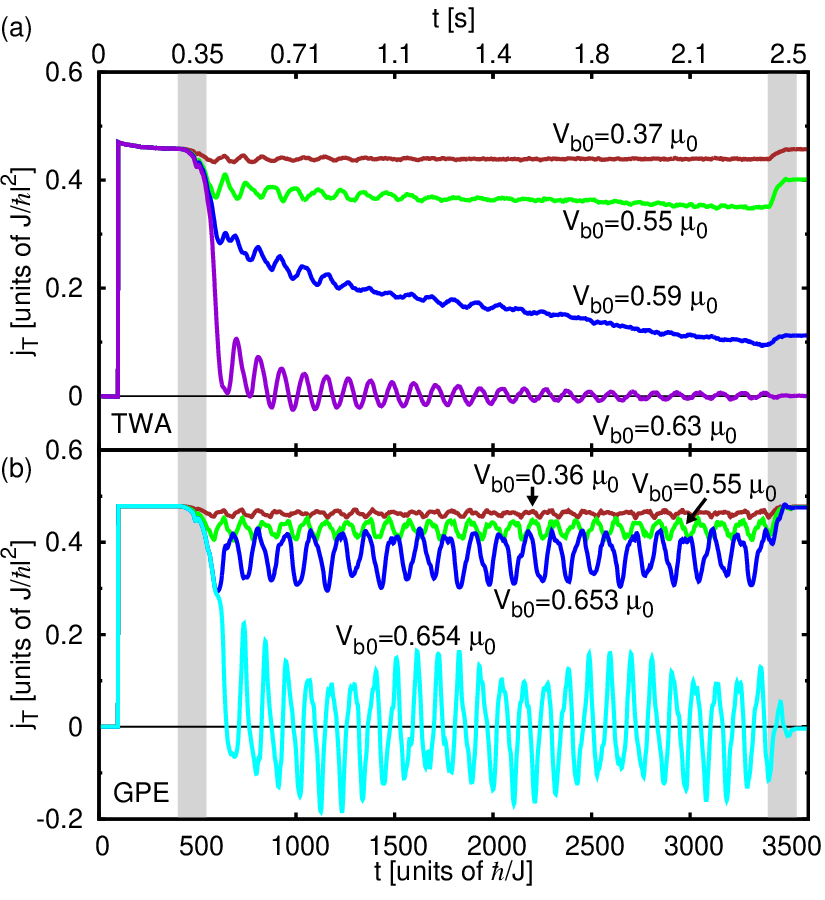}}
\caption{\label{fig:jT_TWA_GPE}(Color online). Time evolution of the average total current density $j_T$ for various barrier heights. The simulations shown here use a  $126 \times 17 \times 5$ lattice, and $N= 51 408$ atoms. The barrier is ramped on and off linearly during the times marked by the shaded regions. Panel (a) shows TWA simulations, which use 512 realizations.  Prior to the barrier ramp-up the chemical potential is $\mu_0 = 2.17J = 0.87\hbar\omega_z$ and the temperature is $T_{TWA} = (4.34 \pm 0.08) J = (46.9 \pm 0.9)$ nK. 
 Panel (b) shows GPE simulations, with $\mu_0=2.20J = 0.88 \hbar\omega_z$.}
\end{figure}
%
In the absence of the barrier potential $V_b(x)$ we find that the superfluid circulates essentially without decay on the simulation time scales, consistent
 with the experimental findings. However, for non-zero barrier heights, decay can occur. In order to characterize this decay, we define the azimuthal component of the average total current density
 $j_T \equiv \left(L_x L_y L_z \right)^{-1}\sum_{\v{r}} j_x(\v{r})$.

 In Fig.~\ref{fig:jT_TWA_GPE} (a) the average total current density, $j_T$, is plotted as a function of time for different barrier heights, for $N=51408$ atoms on a $126 \times 17 \times 5$ lattice.  The barrier height is reported in units of the bulk chemical potential in the absence of the barrier, which is calculated from the density at the trap minimum, averaged over the  azimuthal direction, which for this example is $\mu_0 =2.14 J$ (see Appendix \ref{appendix:chem_potential}).  The barrier is ramped on and off linearly during the time indicated by the gray-shaded regions. The temperature of the TWA simulations is $T=(4.34 \pm 0.08) J =(46.9 \pm 0.9)$ nK, measured before the barrier ramp up, as described in  Appendix \ref{appendix:temperature}.
  For comparison, we also show results from GPE simulations in Fig.~\ref{fig:jT_TWA_GPE} (b). For the GPE simulations, the initial state was generated by using imaginary time propagation to calculate the GPE ground state in the trap \cite{dalfovo_bosons_1996}.
 In the GPE simulations, two scenarios are observed: Either the current decays quickly compared to the experimental time scale, or it persists, and the different  barrier heights only affect the time-averaged value of the current while the barrier is up. For each case an oscillatory behavior is observed that has little damping on the time scale of the experiment. These oscillations are due to excitations generated during the barrier ramp-up and the amplitude of these oscillations decreases as the barrier ramp-up time, $\Delta_{t_b}$, is increased.
 In the TWA simulations the decay behavior crosses over smoothly from small to large barrier heights. For small barriers, the decay is much slower than the experimental times. For larger barrier heights, superfluid decay is visible, as can be seen in the examples for $V_{b0}/\mu_0 = 0.55$ and $V_{b0}/\mu_0 = 0.59$ in Fig.~\ref{fig:jT_TWA_GPE} (a). As the barrier height is further increased, a fast decay is visible, as in the example $V_{b0}/\mu_0 = 0.63$ in Fig.~\ref{fig:jT_TWA_GPE} (a), that is qualitatively similar to the fast decay visible in the GPE simulations. We note however that this decay sets in at smaller values of $V_{b0}$, and thus GPE overestimates the stability of the superfluid flow. We also note that within the TWA simulation, the oscillatory behavior of $j_T$ is damped. 
%
\begin{figure}
\centerline{\includegraphics[scale=0.5]{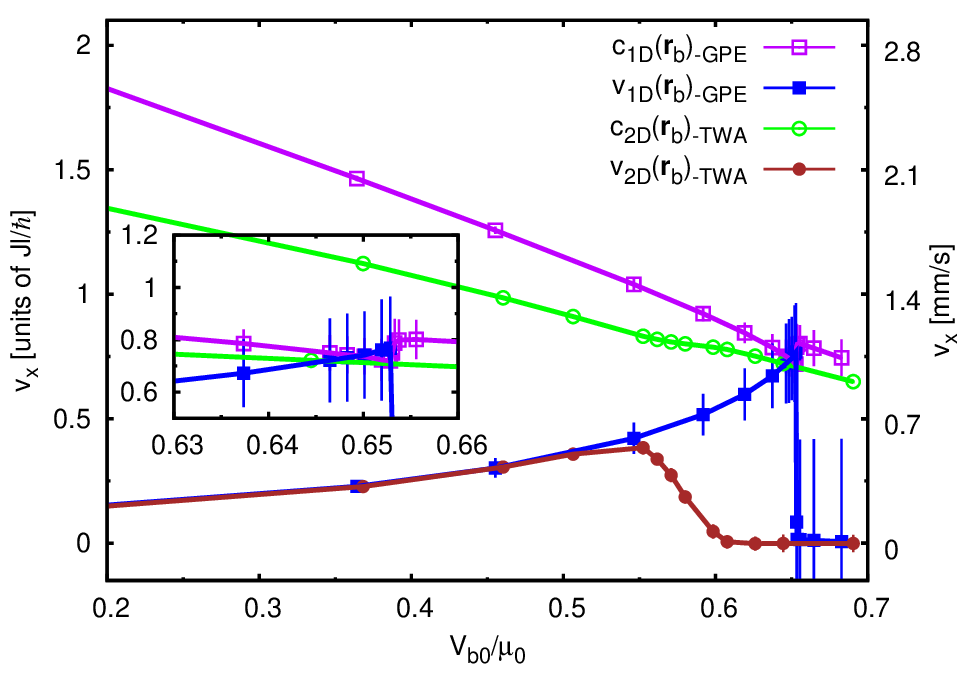}}
\caption{\label{fig:vl_Vb}(Color online). Velocity and speed of sound at the barrier versus barrier height in units of the bulk chemical potential.  GPE data: velocity and speed of sound from the one-dimensional (1D) integrated density and current.  TWA data: velocity and speed of sound from the column density and column current (see Appendix B). Vertical lines indicate the standard deviation of the velocity and sound speed. Inset: enlarged view, close to the GPE critical barrier height. The simulation parameters of the TWA data are the same as for Fig. \ref{fig:jT_TWA_GPE}: The atom number is $N=51408$, the chemical potential $\mu_0 = 2.17 J =0.87\hbar\omega_z$, and the temperature $T_{TWA} = (4.34 \pm 0.08) J = (46.9\pm 0.9)$ nK. We use a  $126\times 17\times 5$ lattice. Note that the 2D (1D) velocity and sound speeds are only strictly valid around the critical barrier heights for the TWA (GPE) data.}
\end{figure}

In Ref.~\cite{ramanathan_superflow_2011} an experimentally motivated definition of the critical velocity of the superfluid was introduced: After the hold time
 of the barrier of typically $2$s, the barrier was removed and the atomic cloud was allowed to expand. From the time-of-flight images it was deduced 
 whether there was phase winding 1, indicating persisting superflow or phase winding 0, indicating decay of the superflow.  For a given barrier height and chemical potential, these events occur probabilistically. 
The experiments defined a critical barrier height at which the probability for the superfluid to decay in the 2 second hold time was $50 \%$.   
 The critical velocity was the calculated velocity of the flow at the barrier for that barrier height. Phrased differently, the velocity was called critical if the superfluid decay time equaled the hold time of the experiment. 
 A full comparison to the experiment will be given in Sect.~\ref{sec:expt}, where we imitate the experimental procedure.

Here, as a first comparison, we calculate the velocity at the barrier maximum at the end of the hold time for the TWA simulations.
 The velocity at the barrier maximum is defined as 
\begin{equation}
v_{2D} \equiv \frac{j_{x\text{-}2D}({\bf r}_b)}{\sqrt{n_{2D}({\bf r}_b)n_{2D}({\bf r}_b+l\hat{\v{x}} ) }}\label{eqn:v2D}
\end{equation}
where ${\bf r}_b\equiv (x_b,0)$, $j_{x\text{-}2D}(\v{r}_b) = \sum_z j_{x} (x_b,0,z)$ is the column current and $n_{2D}(\v{r}_b) =\sum_z n (x_b,0,z)$  is the column density at $(x_b,0)$.
 We note that $j_x({\bf r}_b)$ describes the current along the bond $\langle {\bf r}_b, {\bf r}_b +l\v{\hat{x}}\rangle$.
 In Fig.~\ref{fig:vl_Vb}, this velocity is plotted as a function of the barrier height in units of $\mu_0$. 
 For the TWA simulations, $v_{2D}$ is averaged over a time window $t \in [ 3250, 3395] \hbar/J$ immediately prior to ramping down the barrier, which is a good measure of the velocities just before the time-of-flight measurement in the experiment.  
 By comparing the `local' chemical potential, $\mu(x_b)$ (see Appendix B) to the confining energies $\hbar \omega_y$ and $\hbar \omega_z$, we find
that the dynamics in the $z$-direction is frozen out at the barrier, while the Thomas-Fermi approximation holds for the radial direction, which is why we choose to plot $v_{2D}$. With these assumptions the phonon velocity can be calculated, and expressed in terms of the peak density, leading to the approximate local speed of sound \cite{zaremba_sound_1998, stringari_dynamics_1998},
\begin{eqnarray}
c_{2D}(x_b) & =  & \sqrt{2\mu_{2D}(x_b)/3m}\label{eqn:c2D}
\end{eqnarray}
where $\mu_{2D}(x_b)$ is the chemical potential for a quasi-two-dimensional (2D) system (see Appendices \ref{appendix:chem_potential} and \ref{appendix:sound_speed}).  The local speed of sound is also plotted in Fig.~\ref{fig:vl_Vb}. We compare the local velocity to the local sound speed at the barrier maximum, rather than the edge of the condensate, because the excitations which cause the phase slip must transverse the barrier radially, passing through the center, as will be discussed in more detail in Sect.~IV.

In the GPE simulations, the dynamics at the barrier in both $y$ and $z$ direction are frozen out, making the system locally quasi-one-dimensional (1D), when the barrier height $V_{b0}$ approaches the critical barrier height. Because the system is in the single-channel regime at the barrier, we plot the 1D velocity 
\begin{equation}
v_{1D} \equiv \frac{j_{x\text{-}1D}({\bf r}_b)}{\sqrt{n_{1D}({\bf r}_b)n_{1D}({\bf r}_b+l\hat{\v{x}} ) }}\label{eqn:v1D}
\end{equation}
based on the integrated 1D current, $j_{x\text{-}1D}(\v{r}_b) = \sum_{y,z} j_{x} (x_b,y,z)$ and the integrated 1D density, $n_{1D}(\v{r}_b) = \sum_{y,z} n (x_b,y,z)$. 
The 1D speed of sound at the barrier is also plotted,
\begin{eqnarray}
c_{1D}(x_b) & =  & \sqrt{\mu_{1D}(x_b)/m},\label{eqn:c1D}
\end{eqnarray}
where $\mu_{1D}(x_b)$ is the local chemical potential for a quasi-1D system (see Appendices \ref{appendix:chem_potential} and \ref{appendix:sound_speed}).
For the GPE simulations, the velocity is averaged over a window in the range $t \in [545, 3395]\hbar/J$, during which the barrier was held at its maximum value. As noted before, in the GPE simulations, no sizable decay was observed during the hold time, allowing for this long time interval to be used for averaging. 
The vertical bars represent the standard deviation of the barrier velocity and local speed of sound, which indicates the amplitude of oscillations occurring in these quantities.  These are quite large for the GPE simulations and are due to the undamped oscillations generated during the barrier ramp-up seen in Fig.~\ref{fig:n_jx_sketch}.  

For the TWA (GPE) simulations, the system is quasi-2D (1D) at the barrier only when the barrier height is close to the critical barrier height, but we choose to plot 2D (1D) quantities for all barrier heights because we are most interested in the critical region. 
We note that at the critical barrier height, the healing length at the barrier center ranges from $\xi \approx 1.7-2.6 l$, which is shorter than the total width of the barrier $2 l_b = 6 l$. We thus expect that the system is still consistent with a local-density approximation. 

We recognize in Fig.~\ref{fig:vl_Vb} the features of the TWA and the GPE simulations found before: 
 As the barrier is increased, for both simulations $v_b$ increases gradually, and then falls off to zero as the superfluid flow becomes unstable.
 The transition occurs over a finite range in the TWA simulations, while within GPE, there is a sharp jump between barriers for which the current persists and those for which it decays.
 As can be seen from Fig.~\ref{fig:vl_Vb}, the critical barrier height within TWA is lower than for GPE and the transition from persistent to decaying current
occurs when the local velocity is measurably lower than the local speed of sound, while we find (see also the inset) that within GPE the current decay occurs when the velocity at the barrier is comparable to the local speed of sound, as has been found in other GPE simulations in the hydrodynamic limit \cite{watanabe_critical_2009,piazza_vortex-induced_2009,*piazza_instability_2011}.

\section{Phase slip Dynamics}\label{sec:phase_slip}
%
\begin{figure}
\centerline{\includegraphics[scale=0.5]{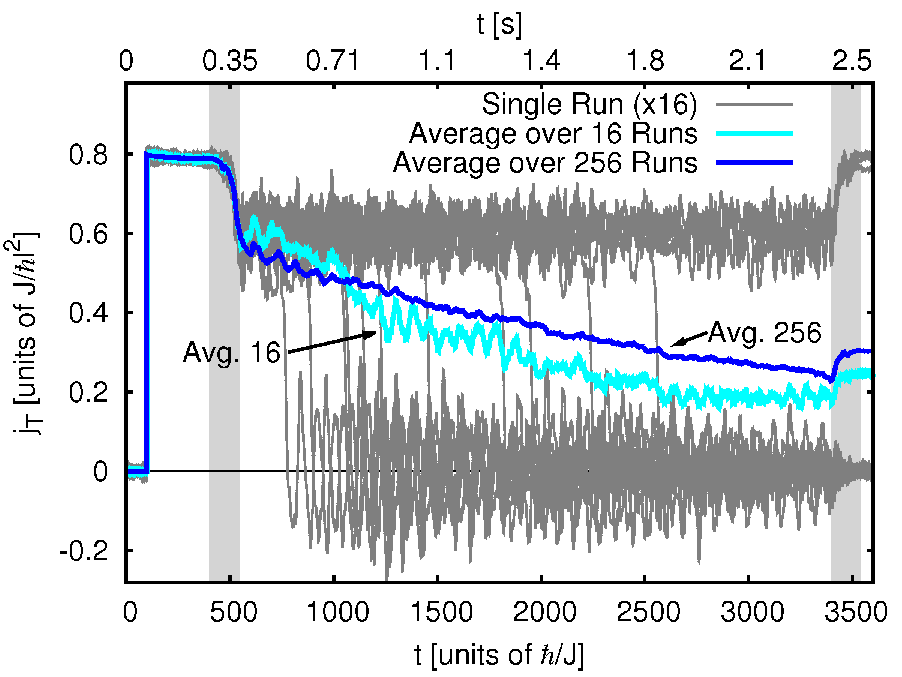}}
\caption{\label{fig:jT_Ipn1}(Color online). Average total current density, $j_T$ for sixteen individual TWA realizations (thin gray lines); $j_T$ averaged over 16 shown realizations and averaged over 256 realizations. The atom number is $N=150028$, the barrier height $V_{b0}=2.67 J= 0.67 \mu_{0}$, and the temperature $T=(5.58 \pm 0.15) J = (60.3 \pm 1.6)$ nK.}
\end{figure}

\begin{figure}
\centerline{\includegraphics[scale=0.5]{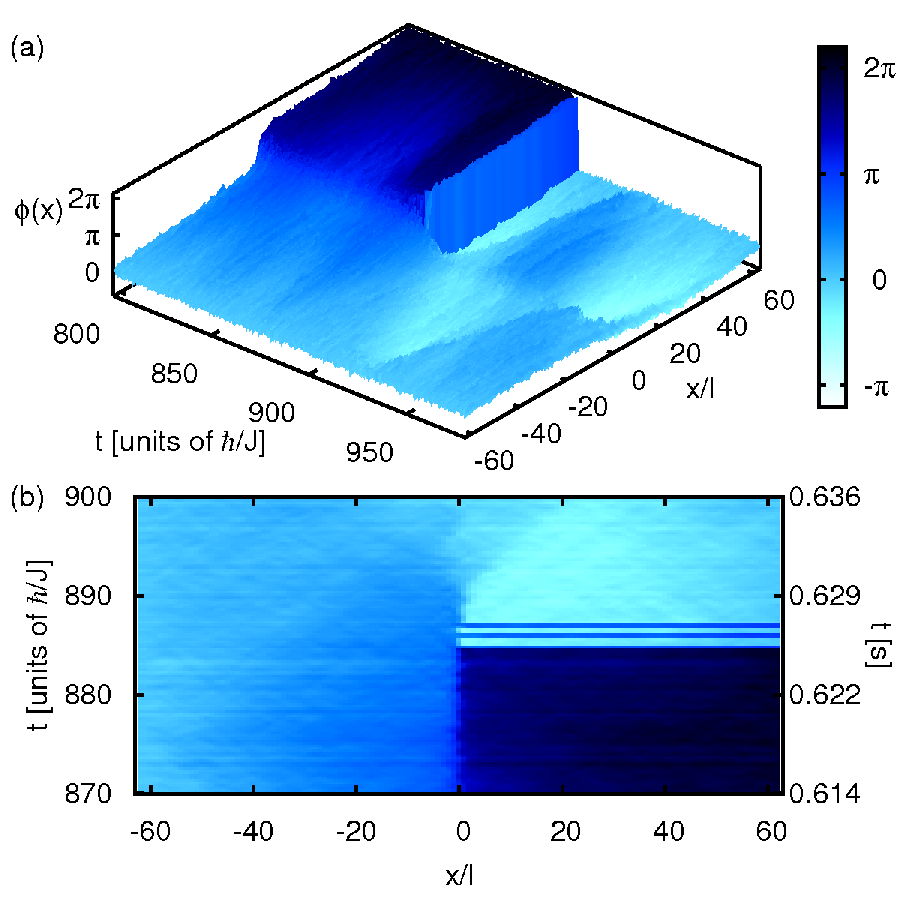}}
\caption{\label{fig:phi_xt}(Color online).  Time evolution of the phase at the center of the ring, $\phi(x) = \phi(x,0,0)-\phi(0,0,0)$ versus $x$ and $t$ for a single TWA realization, with the same parameters as in Fig. \ref{fig:jT_Ipn1}. In Panel (a) we show the time interval $[800,980]\hbar/J$, in Panel (b) the interval   $[870,900]\hbar/J$, as a contour plot. These intervals bracket the phase slip event at $t\approx 885 \hbar/J$.} 
\end{figure}

In order to understand the decay mechanism in the TWA simulations, we investigate the dynamics of the current and phase for individual realizations.
In Fig.~\ref{fig:jT_Ipn1} the average total current density, $j_T$, is plotted for several individual realizations of TWA simulations (gray lines).  Within an individual realization there is a rapid transition from a state with circulation to one without circulation, while the average over many realizations reveals an exponential decay for the ensemble (see also Fig.~\ref{fig:jTtau_Teff_v1}(a)). The time when the decay occurs is probabilistic and is governed by the decay timescale for the ensemble.  To further understand the mechanism of current decay, we look at the phase dynamics of a single realization around the time the total current decays. We define the phase along the density maximum, $\phi(x)=\phi(x,0,0)-\phi(0,0,0)$, where the phase is chosen so that the difference between two neighboring sites is always $\phi(x+l)-\phi(x) \in (-\pi,\pi]$.
The phase along the center of the ring, $\phi(x)$,  is plotted in Fig.~\ref{fig:phi_xt} for a time window after phase imprinting and the barrier ramp-up.  Initially there is a global phase winding of $2\pi$ and the steepest slope of the phase occurs across the barrier, where the density is at a minimum and the velocity at a maximum, due to flow continuity.  Around $t=885 \hbar/J$ the phase at the barrier jumps sharply and the overall phase winding drops to zero. This coincides with the total current dropping to zero.  Subsequently long wavelength excitations are observed in the phase as the system dissipates the energy generated by the phase slip.  

\begin{figure}
\centerline{\includegraphics[scale=0.58]{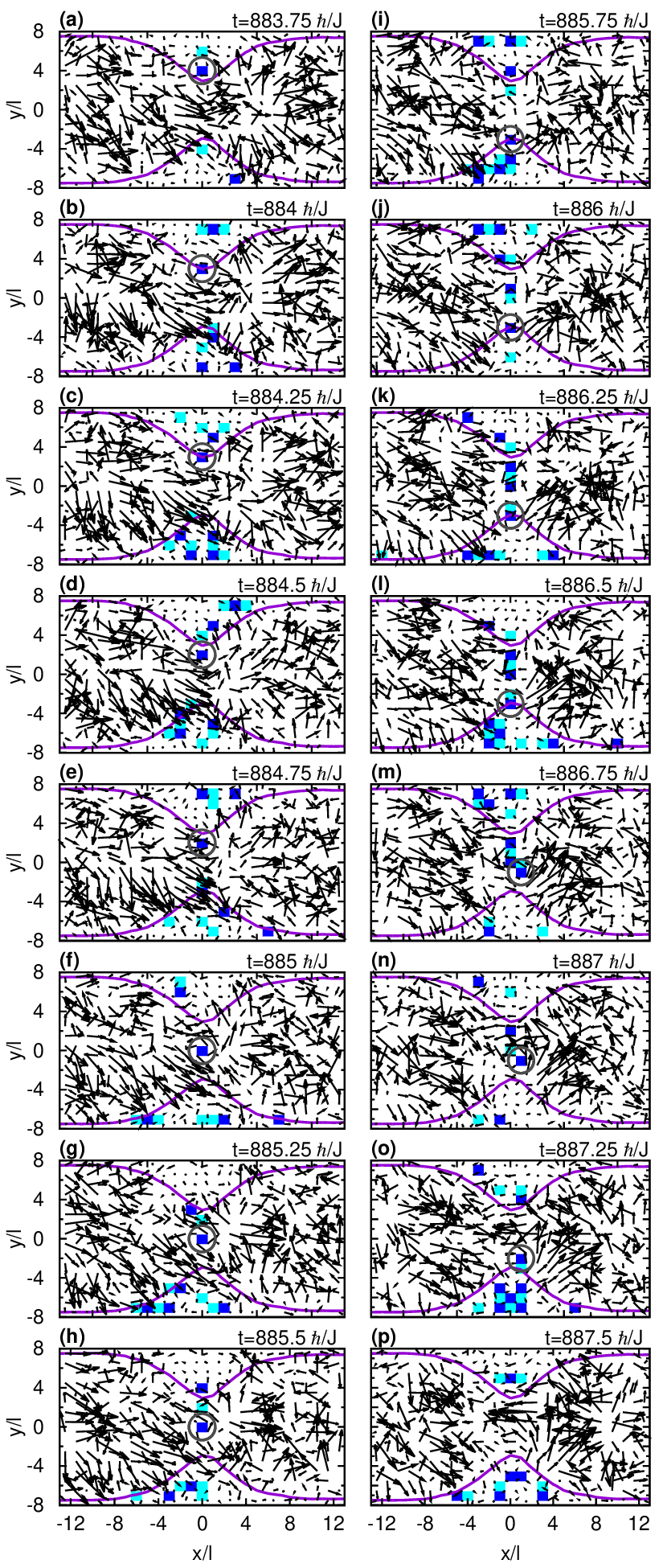}}
\caption{\label{fig:jxy_frames}(Color online). Vortices (light blue (light gray) squares), anti-vortices (dark blue (dark gray) squares) and current $j_{x,y}(x,y,0)$ (vector field), in the barrier region for the same TWA realization shown in Fig.~\ref{fig:phi_xt}. Vortices and anti-vortices are determined from the phase winding around individual plaquettes.  The solid line, which corresponds to an average density that is $10 \%$ of the maximum density, highlights the barrier region.  Data plotted for a cut at $z=0$ for a time sequence from $t=883.75 \hbar/J$, to $t=887.75 \hbar/J$ in increments of $0.25 \hbar/J$.  A single anti-vortex (phase winding $-2\pi$), which is circled, crosses the barrier region, leading to the phase slip.  Vortex-anti-vortex pairs are generated and annihilated in the low density regions. }
\end{figure}
For the same realization, a sequence of snapshots of the vortices, anti-vortices and current field, $j_{x,y}(x,y,0)$, in the barrier region around the time of the phase slip are plotted in Fig.~\ref{fig:jxy_frames}.  The presence of vortices and anti-vortices are calculated from the phase winding around an individual plaquette, $\sum_\square \delta \phi(x,y) = \delta_x\phi(x,y) + \delta_y \phi(x+l,y) - \delta_x \phi(x,y+l) - \delta_y \phi(x,y)$, 
where the phase difference between sites is always taken to be $\delta_x \phi(x,y)\equiv\phi(x+l,y)-\phi(x,y) \in (-\pi,\pi]$. A vortex (anti-vortex) corresponds to a plaquette phase winding of $2\pi (-2\pi)$. Additionally, in order to more clearly identify the barrier region, a solid line traces a constant-density curve $n(x,y,0)$ which is $10\%$ of the maximum density, based on the density time-averaged over an interval of $200\hbar/J$.   

Initially, the current flows to the right (a) and vortex-anti-vortex pairs are created and annihilated in the low density regions at the edges of the ring.  These pairs are seeded by thermal fluctuations and rarely lead to a phase slip \footnote[1]{See online Supplemental Material for a movie of the dynamics of vorticity and current for the entire ring at z = 0. It is clear in the movie that numerous vortex-antivortex pairs are created and annihilated without leading to a global phase slip.}. Around time $884 \hbar/J$, a single anti-vortex penetrates into the barrier region,  which is seen clearly in Fig.~\ref{fig:jxy_frames}(e)-(h). The single anti-vortex is circled to highlight its trajectory although it is not always possible to unambiguously identify this single anti-vortex when several are present.  Vortices and anti-vortices, which are attracted to the single anti-vortex, penetrate into the barrier region from the edges or are generated in the barrier region (i)-(m).  Eventually the additional vortex-anti-vortex pairs in the barrier region annihilate (m)-(o). Finally the single anti-vortex crosses to the other side, the current in the barrier region changes direction and subsequently the total current decays. 
The complex dynamics of vortices and anti-vortices in the barrier region give rise to the oscillations in the total phase winding that occur around the time of the phase slip and are observed in Fig.~\ref{fig:phi_xt}(b). The time for the single vortex to cross barrier region in this example is approximately $5 \hbar/J \approx 3.5$ ms and the decay of the total superflow is observable on a timescale of about $15 \hbar/J \approx 10.5$ ms.

Including the curvature in the simulations would lead to a difference in the average velocity of the inner and outer edges.  However this velocity difference is not significant compared to the thermally induced fluctuations of the velocity and thus is not expected to change the parameters where the phase slip becomes favorable. If the curvature is included in the GPE simulations, we find that the phase slip is always caused by a vortex entering the ring from the inner edge, as was also observed by \cite{piazza_critical_2013}.  In the TWA simulations, without the curvature, it is equally likely that the phase slip is caused by an antivortex entering from above as seen in Fig.\ref{fig:jxy_frames} as that it is caused by a vortex entering from below. If the curvature were included, a preference for phase slips caused by vortices entering the ring from the inner edge might persist to small temperatures, but may not be observable at the temperatures of the simulations reported here.

For the TWA simulations, the nucleation of vortices, which occurs continually in the low-density regions, does not govern the decay of the superfluid flow, but rather the passage of a single vortex or anti-vortex across the barrier.  The correct criterion for superfluid decay therefore cannot be that the cloud is susceptible to vortices penetrating the surface, but rather a stronger criterion has to be used, one that takes into account that the higher densities at the center of the trap have to be transversed as well. For that to happen, one can expect that in the Landau criterion the sound velocity of the peak density should be used.  This supports the choice of the critical velocity at the center of the barrier, rather than the edges, as the relevant measure for the superfluid decay.

In the GPE simulations, the superfluid decay occurs on a much faster time-scale.  We note that for the GPE data depicted in Fig.~3, the system in closest to a 1D system, but not deep in the 1D system, so that some transverse dynamics are still present, and the decay process corresponds to a vortex traveling very quickly through the barrier region.  If we ignore the existence of the weak occupation of higher modes, besides the ground state mode of the harmonic oscillator, i.e. we imagine to project on the lowest mode, the phase slip will indeed look very similar to a soliton in the 1D system. These two types of defects, however, are continually connected in these systems, which are in the regime of dimensional cross-over.

\section{Comparison to Experiment}\label{sec:expt}
In this section we compare the simulations directly to experiment. First, we analyze the relationship between the critical barrier and chemical potential, and second, the critical velocity compared to the speed of sound.
 In order to imitate the data analysis performed in Ref.~\cite{ramanathan_superflow_2011}, we determine the critical barrier height as follows.  For the TWA simulations, we fit the decay of the total current with an exponential function $j_T=j_0 \exp[-(t-t_0)/\tau]$, with fitting parameters $j_0$ and $\tau$, to determine the decay time scale $\tau$. As discussed earlier, the analysis of the experiment determined the critical parameters by finding the conditions under which half of the initial realizations had decayed to zero phase winding. Therefore we define the critical decay time scale as $\tau_{cr} = \Delta t /\log 2$, which corresponds to $50\%$  probability of decay of the current during the experiment. Here $\Delta t = 2850 \hbar/J$ is the time the barrier is held at its maximum value, and we ignored the decay which occurs while the barrier is ramped on and off.  
 We then interpolate the $\tau$-versus-$V_{b0}$ relation with $\tau= \tau_0\exp(\alpha V_{b0})$, and define
  the critical barrier height as $V_{bcr}\equiv V_{b}(\tau_{cr})$.  
  For the GPE simulations, the critical barrier value shown is the average of the largest barrier height with no decay and the smallest barrier height that shows decay.

\begin{figure}
\centerline{\includegraphics[scale=0.55]{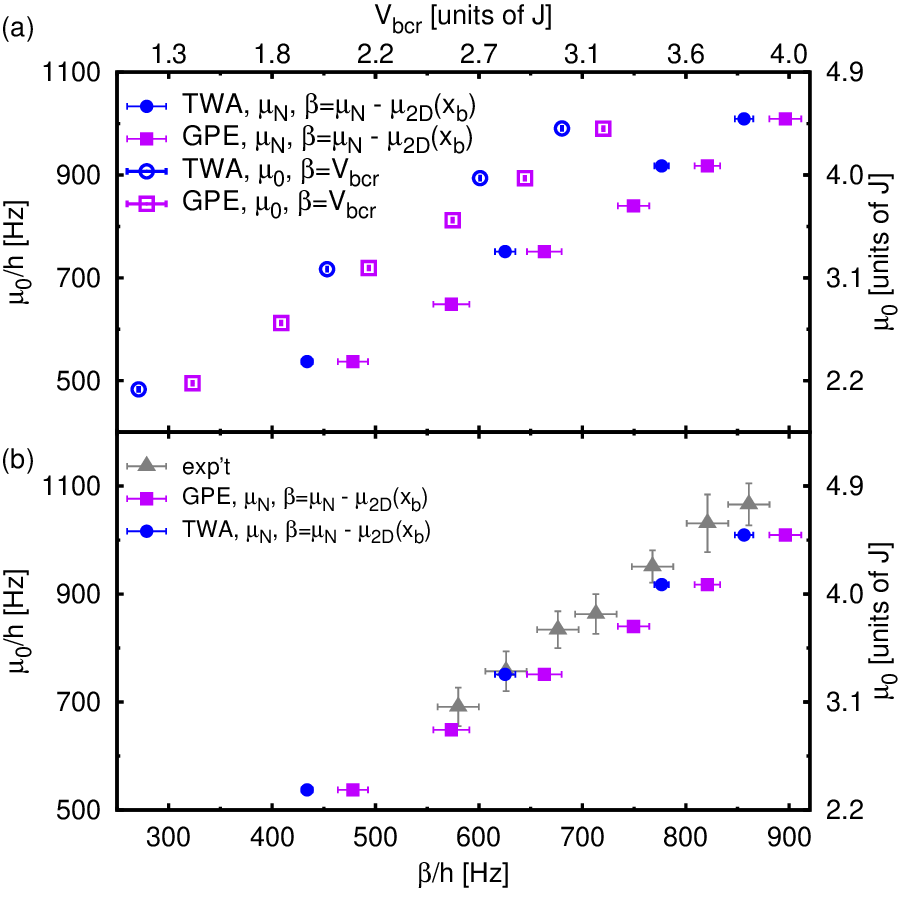}}
\caption{\label{fig:muc_Vb}(Color online). In Panel (a) we show the chemical potential versus the critical barrier height, based on two ways of analyzing the simulation results. The   open symbols depict the chemical potential  $\mu_0$, calculated from the density distribution prior to the barrier ramp-up, and the critical barrier height $\beta= V_{bcr}$, using the barrier potential directly. The solid symbols depict the analysis that resembles the analysis of Ref. \cite{ramanathan_superflow_2011}.
 We use $\mu_{N}$ as the approximation for the chemical potential, calculated from the Thomas-Fermi distribution for $N$ atoms, and $\beta=\mu_{N}-\mu_{2D}(x_b)$ as the approximation of the barrier height, where $\mu_{2D}(x_b)$ is the local chemical potential at the barrier, determined from the column density. The temperature of the TWA simulations are $T_{TWA}=(5.50 \pm 0.13) J = (59.4 \pm 1.4)$ nK. In Panel (b) we show the comparison of the experimental data and the simulation data processed in the same way as the experimental data in Ref. \cite{ramanathan_superflow_2011}. The experimental data are the triangular symbols, the simulation data are the solid, circular and square symbols, which are the same as in Panel (a).}
\end{figure}
In Fig.~\ref{fig:muc_Vb}(a)  we show the density-based approximation for the chemical potential, $\mu_0$, (see Appendix \ref{appendix:chem_potential}) plotted as a function of
 the critical barrier height $V_{bcr}$, depicted by the open symbols for the TWA and the GPE simulations.  
  For fixed chemical potential, the current decays for lower barrier heights when quantum and thermal fluctuations are included, i.e. in the TWA simulations, compared to the GPE simulations.  These findings are consistent with the results shown in Figs.~\ref{fig:jT_TWA_GPE} and \ref{fig:vl_Vb}. 

In the analysis of the experiment,  the chemical potential was calculated from the total number of atoms, assuming a Thomas-Fermi distribution (in the absence of the barrier) in both radial and $z$-directions, $\mu_{N} = \left[ g N m\omega_y \omega_z /(\pi L_x )\right]^{1/2}$, where $N$ is the total number of atoms. The barrier height was approximately determined by $\beta = \mu_{N} - \mu_{2D}(x_b)$ where $\mu_{2D}(x_b)$ is the chemical potential at the barrier maximum, $\mu_{2D}(x_b) = \sqrt{m\omega_z/(2\pi\hbar)}\,gn_{2D}(x_b,0)$, determined from the local column density, $n_{2D}(x_b,0) $  \footnote[2]{We corrected for a missing factor of  $1/\sqrt{2}$ in $\mu_{2D}$ in Ref.~\cite{ramanathan_superflow_2011}.}. 
As discussed in Appendix \ref{appendix:chem_potential}, this expression for the chemical potential assumes that at the barrier only the harmonic oscillator ground state is occupied in the vertical direction, which is valid when $\mu(x_b) < \hbar\omega_z$, as is the case for all of the data presented here. 

In order to compare our results with the experimental results, we generate quantities similar to those studied in the experiment.  In Fig.~\ref{fig:muc_Vb}(a) we plot the chemical potential estimate $\mu_{N}$ and the approximate critical barrier $\beta=\mu_{N} - \mu_{2D}(x_b)$ as approximate quantities based on the TWA and GPE data, depicted by solid symbols, in comparison to the results for $\mu_0$ and $V_{bcr}$. We note that the approximations used in the original experimental analysis overestimate the critical barrier potential.  We account for this discrepancy in Fig.~\ref{fig:muc_Vb}(b), where the simulation results for the chemical potential estimate $\mu_{N}$ and the approximate critical barrier $\beta=\mu_{N} - \mu_{2D}(x_b)$ are re-plotted, along with the experimental data. In these plots, we see that the experimental results are close to both the TWA and GPE simulations.
One can speculate that the GPE simulations predict decay at barriers larger than those observed in the experiment, while the TWA simulations are closer to the experimental results, but the difference is not significant compared to the experimental error bars. 

\begin{figure}
\centerline{\includegraphics[scale=0.6]{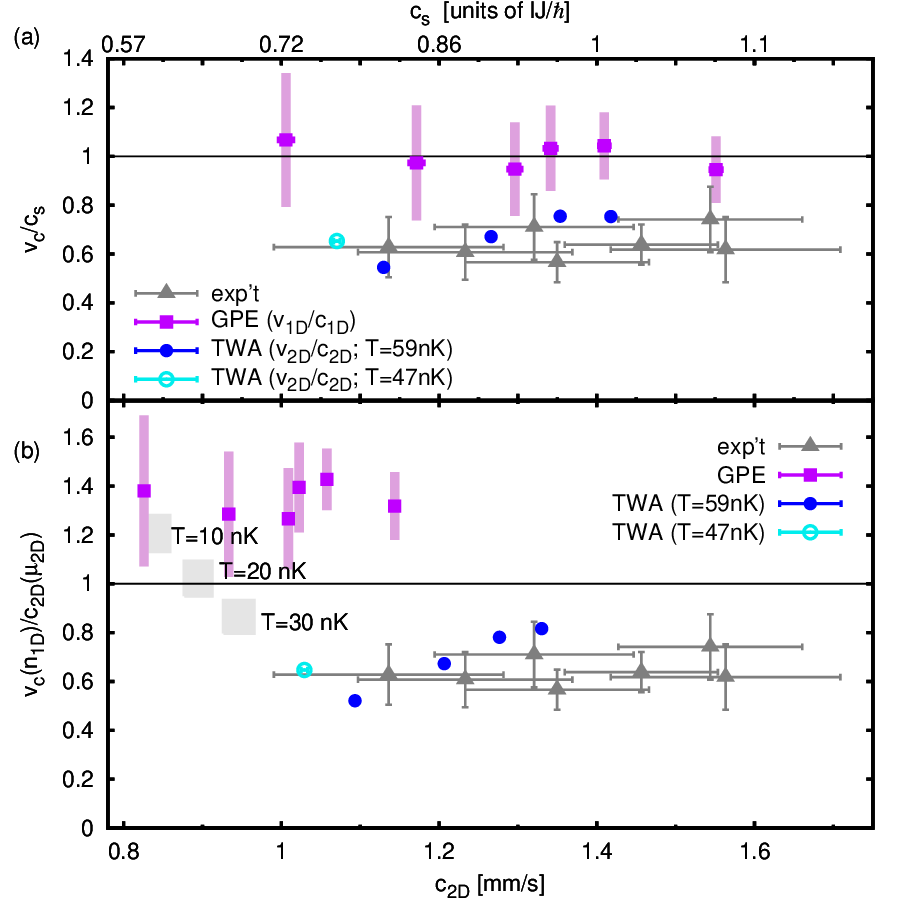}}
\caption{\label{fig:vc_cs} (Color online). Critical velocity versus local speed of sound.  (a) TWA data: $v_{2D}/c_{2D}$ vs. $c_{2D}$, for two different temperatures. GPE data: $v_{1D}/c_{1D}$ vs. $c_{1D}$.  The experimental data (from \cite{ramanathan_superflow_2011}, triangular data points) is rescaled to $c_{2D}$. (b) $v_{2D}/c_{2D}$ for both TWA and GPE data, where $v_{2D}$ is calculated from the 1D integrated density as was done in the experiment.  Experimental data same as in (a).   The gray shaded region represents the estimate of the critical velocity for temperatures in the range of $10$, $20$ and $30$nK, as described in the text.
The experimental data is inconsistent with $v_c / c_{2D} = 1$. Thermally activated phase slips offer a possible explanation of the reduction of $v_c / c_{2D}$ below $1$.}
\end{figure}

However, we do find significant differences for the critical velocity, which we present now.

One finding of the experiment was that the critical velocity - as defined in Ref.~\cite{ramanathan_superflow_2011} - is less than the speed of sound at the barrier maximum, which was determined from the column density. 
 In Fig.~{\ref{fig:vc_cs}} we show a re-analyzed version of the data of Ref.~\cite{ramanathan_superflow_2011}; 
 in particular, the critical velocity, normalized by the local speed of sound, is plotted as a function of the local speed of sound
 \footnote[3]{In  Ref.~\cite{ramanathan_superflow_2011}, the 3D speed of sound for a cylindrical BEC, $c_{3D} = \sqrt{\mu/2m}$, was used, which is valid in the limit of a Thomas-Fermi distribution in both $y$ and $z$ direction.  Here, we have plotted the experimental data in terms of $c_{2D} = \sqrt{2\mu_{2D}(x_b)/3m}$ (see Appendix \ref{appendix:sound_speed}),  where we adjusted for the factor of $1/\sqrt{2}$ missing in $\mu_{2D}$ as in \cite{Note2}. This leads to an overall minor rescaling of the experimental data by a factor of $\sqrt{2\sqrt{2}/3}\approx 0.97$}. We calculate the ratio of the critical velocity to the local sound speed in  two different ways.  First we calculate the ratio based on the local velocity and density, adjusting for the effective dimensionality at the barrier, and secondly we calculate the velocity from the integrated 1D density profile and use the 2D local sound speed, as was done in the experiment, plotted in Figs.~\ref{fig:vc_cs}(a) and \ref{fig:vc_cs}(b), respectively.

In Fig.~\ref{fig:vc_cs}(a) the critical velocity for the TWA simulations was calculated as follows. For a given critical barrier height, $V_\text{bcr}$, and chemical potential, $\mu_0$,  all of the realizations were divided into two groups at each time step: those in which unit phase winding persisted, and those in which the current had decayed.  The persistence or decay of the current was determined from the phase winding around the ring, $\phi(L_x,0,0) - \phi(0,0,0) = \sum_{x=0}^{x=L_x} \sin ^{-1}\left(  \frac{v_x(x)}{2Jl}\right)$, which was calculated at each time step. Subsequently the 2D velocity  at the barrier was calculated directly using Eqn.~(\ref{eqn:v2D}) and averaging only over the realizations in which the global current remained.  The critical velocity is compared with the 2D sound density \footnote[4]{Because the velocity is calculated on the bond from $ \protect \langle  \mathbf{r}_b, \mathbf{r}_b+l\mathbf{\hat{x}} \protect \rangle$, for the data  in Fig.~\ref{fig:vc_cs}(a), the local speed of sound is calculated by averaging over sites $x_b$ and $x_b+l$:  $c_{2D} = \sqrt{\left[\mu_{2D}(x_b) + \mu_{2D}(x_b+l)\right]/3m }$  and  $c_{1D} = \sqrt{\left[\mu_{1D}(x_b) + \mu_{1D}(x_b+l)\right]/2m }$}, because for all of the TWA data, the chemical potential at the barrier satisfies $\hbar \omega_y < \mu_{2D}(x_b) < \hbar \omega_z$ at the critical barrier height.


For the GPE simulations, we plot the critical 1D velocity based on the time-averaged current and density integrated radially, $v_{1D}$, as given by Eqn.~(\ref{eqn:v1D}), compared with the 1D speed of sound, because the system is effectively 1D at the critical barrier height for the GPE data based on the local chemical potential, $\mu_{1D}(x_b)$ \cite{Note4}.
The shaded bars on the GPE data in Fig.~\ref{fig:vc_cs} represent the magnitude of oscillation of the velocity during the dynamics, which are due to the barrier ramp-up. For the TWA simulations,  data is presented with  $T=5.50J = 59$  nK,  as well as one data point with $T=4.34J = 47$ nK. Due to the heating that occurs in the initialization scheme used for the TWA simulations, data sets with lower temperatures were not generated. 

For comparison with the experimental data, the velocity at the critical barrier was also calculated from the 1D density profile and the continuity equation as was done in Ref.~\cite{ramanathan_superflow_2011}. Assuming a steady-state flow, $v_c(n_{1D}) = J_0/n_{1D}(x_b)$, where $J_0 = \frac{2 \pi \hbar}{m}[ \oint dx/n_{1D}(x) ]^{-1}$.  Both the TWA data and GPE data are compared with the 2D speed of sound (Eqn.~\ref{eqn:c2D}). This comparison is plotted in Fig.~\ref{fig:vc_cs}(b). The primary difference in the GPE data in Fig.~\ref{fig:vc_cs}(a) and (b) is the comparison with $c_{2D}$ rather than $c_{1D}$, while the difference in velocity between the two calculations is less than $2\%$.


Interestingly, the comparison between the experimental data and the simulations, shown in Fig.~{\ref{fig:vc_cs}}, demonstrates that the experimental data is not consistent with the GPE simulation. In the GPE simulation, a critical velocity of the magnitude of the local phonon velocity is predicted, when the critical velocity is compared with the 1D speed of sound, which is the most relevant physical quantity.   We note that the critical chemical potential is always in the 2D regime in the experimental data.

On the other hand, the TWA simulations offer a possible explanation: Thermally activated phase slips, which can also be visualized as vortices passing through the barrier, lead to a reduction of the critical velocity below the phonon velocity. 
The TWA simulations at $59$ nK  and the single data point at $47$ nK are within the range of the experimental error bars, however, in the experiment, the temperature of the cloud was reported to be of the order of 10 nK. This was an estimate of the temperature in the absence of heating caused by the stirring beam; considering the uncertainty of this estimate and additional heating due to the stirring beam, actual temperatures of tens of nK cannot be ruled out \footnote[5]{Private communication with the authors of Ref. \cite{ramanathan_superflow_2011}}.  The shaded regions represent an estimate of the critical velocity for temperatures in the range of $10$, $20$ and $30$nK, based on extrapolating the TWA results for $\mu_0=2.14 J$, as we discuss in the next section. The extrapolation to lower temperatures is complicated by the change in dimensionality at the barrier that is expected to occur at lower temperatures, based on the GPE simulations. The comparison in Fig.~\ref{fig:vc_cs}(b) suggests that in addition to thermal fluctuations, other  mechanisms are present in Ref.~\cite{ramanathan_superflow_2011}, which are not captured in the simulations and lead to a further reduction of the critical velocity.  Possible mechanisms include the visible disorder of the trapping potential and technical noise.

\section{Temperature dependence of Superfluid Decay}\label{sec:temp}
%
\begin{figure}
\centerline{\includegraphics[scale=0.55]{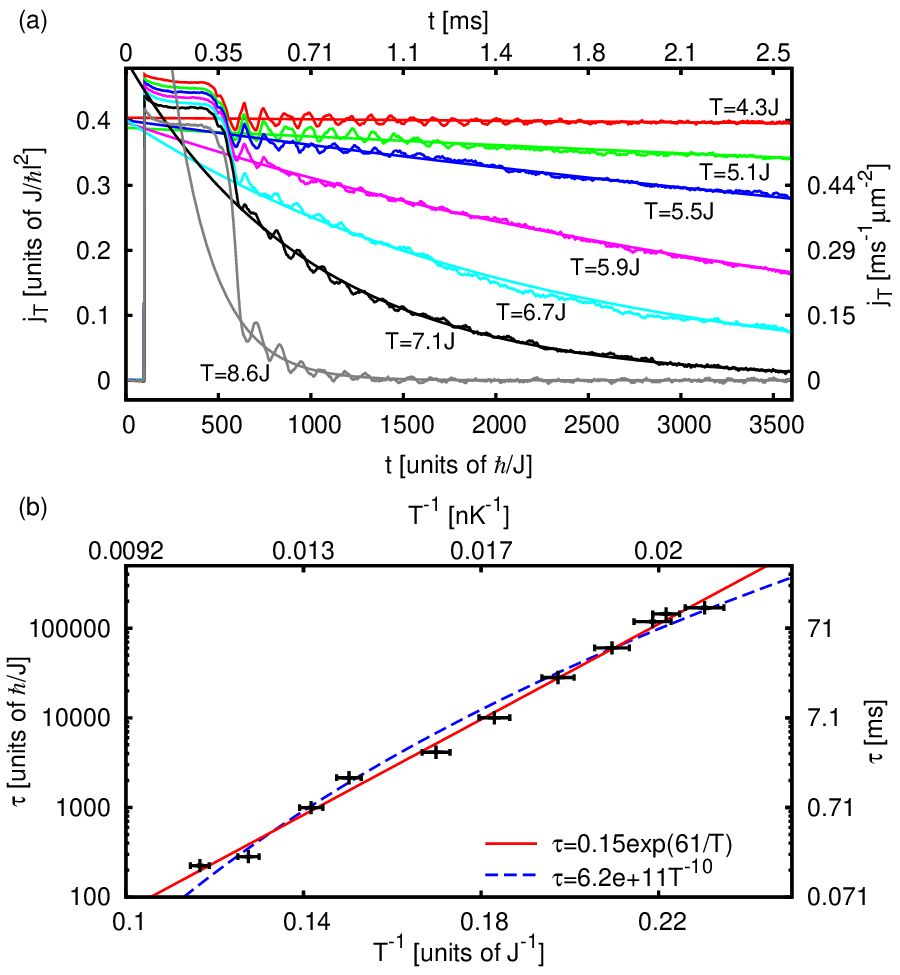}}
\caption{\label{fig:jTtau_Teff_v1} (Color online). (a) Average total current density $j_T$ versus time for different initial temperatures and numerical fits with $A\exp(-t/\tau)$ for fixed barrier height $V_{b0}=1.16J = 0.49\mu_0$.  The total atom number is $N=51407 $ atoms, the lattice dimensions are $126 \times 17 \times 5$. (b) Time scale $\tau$ of the  
current decay versus inverse temperature on a log-linear scale. Numerical fits yield $\tau_j = (0.155 \pm 0.046)\exp\left[(61.3\pm 1.6)/T\right]$ and $\tau_j=(6.23 \pm 4.52)\times 10^{11}T^{-10.3\pm0.41}$. } 
\end{figure}	
%
In this section we study the temperature dependence of the superfluid decay. We keep the system parameters, such as the barrier height, fixed and vary the temperature only.
In Fig.~\ref{fig:jTtau_Teff_v1}(a)  we plot the total current as a function of time for different temperatures, with the barrier height fixed to $V_{b0}=1.16J =0.49\mu_0$. 
 The total current is then fitted with the function $j_T=A\exp(-t/\tau_j)$, over the time window that the barrier was at full power, to determine the decay time scale $\tau_j$.
 In Fig.~\ref{fig:jTtau_Teff_v1}(b), we plot this time scale as a function of the inverse temperature on a log-linear scale.  
  We observe a strong temperature dependence.
   
 To quantify this, we attempt to fit the data with both a power law and an exponential function.
  The exponential scaling  is motivated by theoretical work on superfluids \cite{langer_intrinsic_1967-1} and thin superconducting wires \cite{langer_intrinsic_1967, *mccumber_time_1970}, which found that  the timescale for a $2\pi$ decrease in the phase winding has an exponential dependence, $\tau \sim \exp\left( \Delta F/k_BT\right)$, where $ \Delta F$ is the minimum free energy barrier between the two states. 
  The algebraic scaling is motivated by the behavior of 1D superfluids, see e.g.~\cite{giamarchi_quantum_2004}. We note that at the barrier the system is close to quasi 1D, because the mean-field energy is smaller than $\hbar\omega_z$ and comparable to $\hbar\omega_y$.
 As we see in Fig.~\ref{fig:jTtau_Teff_v1}(b), the data is consistent both with exponential scaling as well as with power-law scaling.  It is not possible to distinguish between the two because the temperature range that is accessible experimentally and in the simulations is rather narrow. 
 We also note that the physical setup might not give rise to pure exponential or algebraic behavior, because phase slips can both occur directly in the region at the barrier that is close to quasi-1D and also slighty away from that region.  However, the strong temperature dependence of the decay timescale indicates the importance of thermally activated processes, which should be measurable in experiments.
 
\begin{figure}
\centerline{\includegraphics[scale=0.5]{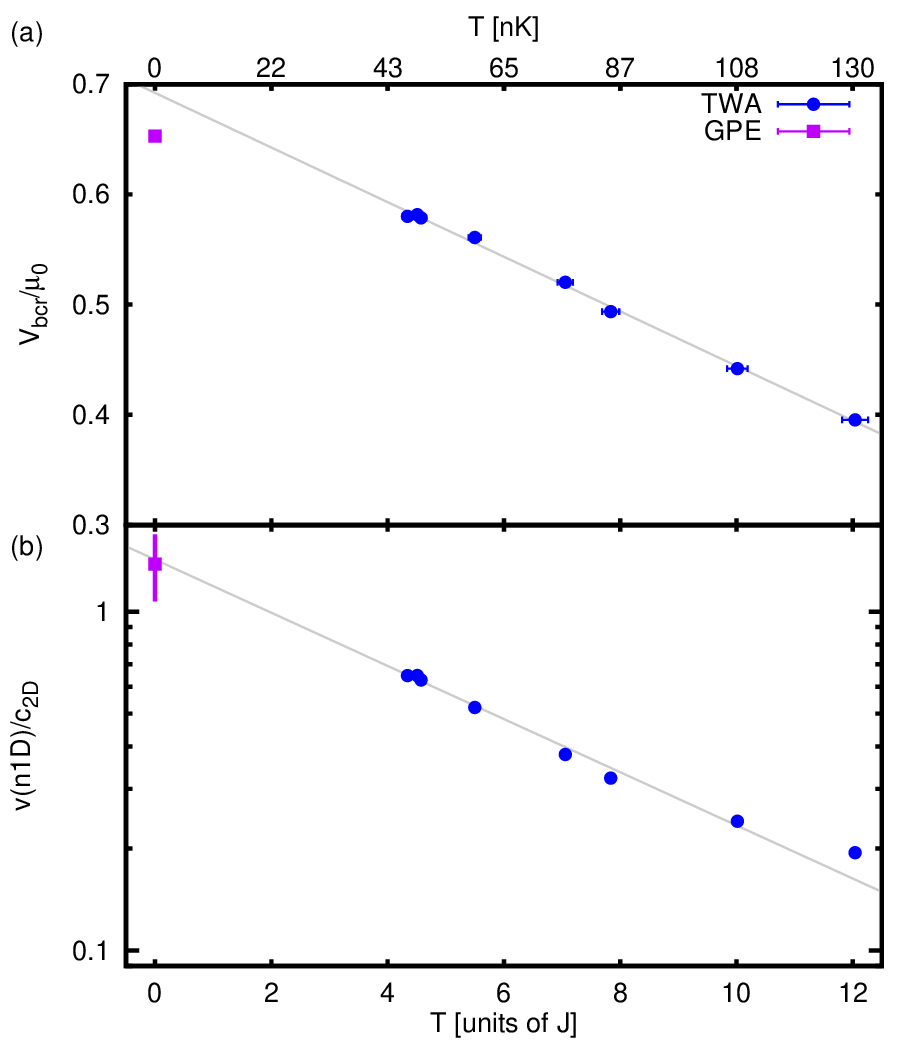}}
\caption{\label{fig:vc_Teff} (Color online). (a) Critical barrier height normalized by the chemical potential, as a function of the initial temperature for TWA (circles) and GPE (square) simulations. A linear fit to the TWA data only gives $V_{bcr}/\mu_0= (0.692 \pm 0.004) -(2.48 \pm 0.05)\times 10^{-2} T/J $. (b) Critical velocity normalized by the local speed of sound at the center of the barrier, as a function of the initial temperature.  An exponential fit to the TWA data only gives $ v_c/c_{2D} = (1.42 \pm 0.07)\exp\left[-(0.181 \pm 0.009) T/J\right]$. This fit is used to determine $v_c/c_{2D}$ for the temperatures $10$, $20$ and $30$nK, shown in Fig.\ref{fig:vc_cs}(b). The lattice dimensions are $126 \times 17 \times 5$ and the total atom number is N = 51407 atoms.  The chemical potential in the absence of the barrier is $\mu_0 = 2.14J$.} 

\end{figure}

Next, we investigate the temperature dependence of the critical barrier and velocity, while keeping the total number of atoms fixed.  In Fig.~\ref{fig:vc_Teff} we plot (a) the critical barrier and (b) the critical velocity as a function of the temperature prior to phase imprinting. In addition, we plot the GPE prediction at $T=0$. We note that the GPE approximation ignores quantum fluctuations, and is thus not the actual $T=0$ prediction.
 For the range of temperatures simulated, the critical barrier height depends approximately linearly on the temperature, as seen in Fig.~\ref{fig:vc_Teff}(a). The gray line is a linear fit to the TWA data only and overestimates the zero temperature critical barrier compared with the critical barrier height from the GPE simulations. The nonlinear dependence on temperature is likely influenced by the changing dimensionality at the barrier that occurs as the temperature is lowered: the dimensionality is effectively 1D at the critical barrier in the GPE simulations, but 2D at the critical barrier in the finite-temperature TWA simulations.

The ratio of the critical velocity to the local sound speed, $v_c(n_{1D})/c_{2D}$, calculated from the 1D density profile, as a function of temperature is plotted on a log-linear scale in Fig.~\ref{fig:vc_Teff}(b) and suggest an  approximately exponential relationship.  Fitting only the TWA data to an exponential function yields a fit $v_c/c_{2D} = (1.42 \pm 0.07)\exp\left[-(0.181 \pm 0.009) T/J\right]$, which agrees with the GPE result at zero temperature, calculated in a similar manner. The 2D speed of sound, calculated from the local chemical potential (Appendix \ref{appendix:sound_speed}) was found to depend on temperature as $c_{2D}(x_b) = (0.040\pm 0.001)T/J + (0.57\pm0.01)$.  We use these fits to estimate the critical velocity at temperatures closer to those in the experiment in Fig~\ref{fig:vc_cs}(b).  The data is complicated by the changing dimensionality at the barrier as the temperature is lowered so that it is not possible to accurately predict the critical velocity at the experimentally relevant teperatures, based soley on temperature range explored here.


We thus find that taking into account thermal fluctuations, within a TWA approach, gives a reduction of $v_c/c_{2D}$ from the GPE predictions, but less than that found in experiment.  Further contributions to the decay that are present in the experiment but are not captured in the simulations could be disorder in the trapping potential or other technical noise, which would lower $v_c/c_{2D}$ further.  

\section{Conclusion}\label{sec:conclusion}
%
In conclusion, we have simulated the experiment reported in  \cite{ramanathan_superflow_2011}, using a TWA simulation and, for comparison, a GPE approach. 
 We find that thermal fluctuations captured within TWA simulations significantly modify the results of GPE simulations. In particular, the critical barrier height and the critical velocity  -- as defined in \cite{ramanathan_superflow_2011} -- are reduced.
 Furthermore, by observing individual TWA realizations, we identify the decay  mechanism of superfluid flow in a toroidal BEC to be thermally activated phase slips at the barrier, which are generated by vortices crossing the barrier region. We also study the temperature dependence of the decay time scale, and find a strong dependence, as shown in Fig.~\ref{fig:jTtau_Teff_v1}. This dependence could be used to experimentally verify thermally activated phase slips at the decay mechanism.

 We compare our results with the experimental results reported in \cite{ramanathan_superflow_2011}, as shown in Figs.~\ref{fig:muc_Vb} and \ref{fig:vc_cs}. 
  These experiments had found that $v_c/c_{eff} \approx 0.55-0.85$. We find this to be in contradiction to GPE simulations which give approximately $v_c/c_{eff} \approx 1$.
  However, taking into account thermal fluctuations within a TWA simulation offers a possible explanation. For temperatures higher than those in the experiments, we find a reduction of the critical velocity comparable to that seen in the experiments. Thus the reduction that was found in experiment appears to be even larger than the thermal reduction. This suggests that besides the thermal effects that are simulated here, additional effects such as the visible disorder of the trap potential could reduce the critical velocity to the experimentally observed regime. This emphasizes the importance of including fluctuations in the simulations of ultra-cold atom systems, to understand `post-GPE' dynamics.   

\acknowledgments
We thank K. C. Wright,  A. Ramanathan, C. J. Lobb, W. D. Phillips, and G. K. Campbell for the experimental data and for invaluable discussions on the experimental methods. This work was supported by the NSF under Physics Frontier Grant No. PHY-0822671. We acknowledge support from the Deutsche Forschungsgemeinschaft through the SFB 925 and the Hamburg Centre for Ultrafast Imaging, and from the Landesexzellenzinitiative Hamburg, which is supported by the Joachim Herz Stiftung.  ACM additionally acknowledges support from NRC/NIST. 

\appendix

\section{Measuring Temperature via coupling and decoupling harmonic oscillators}\label{appendix:temperature}
%
In order to investigate the temperature dependence of the superfluid decay we developed a method to measure the temperature of the atomic cloud in the trap.
 We use this method prior to the phase imprint, as the last step of the preparation stage of the numerical simulation.    

We first couple several harmonic oscillators weakly and adiabatically to the $x$- or $y$-component of the current, $j_{x,y}({\bf r}_s)$, at different locations ${\bf r}_s$ in the ring.  The harmonic oscillator Hamiltonian takes the form:
\begin{align*}
H_{ho} =  \sum_s H_{0s}  + U_{ho}\gamma(t_n) \sum_s p_s j_{\gv{\alpha}} (\v{r_s})
\end{align*}
where $H_{0s} = \frac{1}{2} \left ( p_s^2 + \omega_{ho}^2 x_s^2 \right)$ is the bare harmonic oscillator  Hamiltonian and $x_s$ and $p_s$ are the position and momentum of oscillator $s$, where $s = 1, \ldots ,N_{ho}$ and $N_{ho}$ is the number of oscillators.

\begin{figure}
\centerline{\includegraphics[scale=0.5]{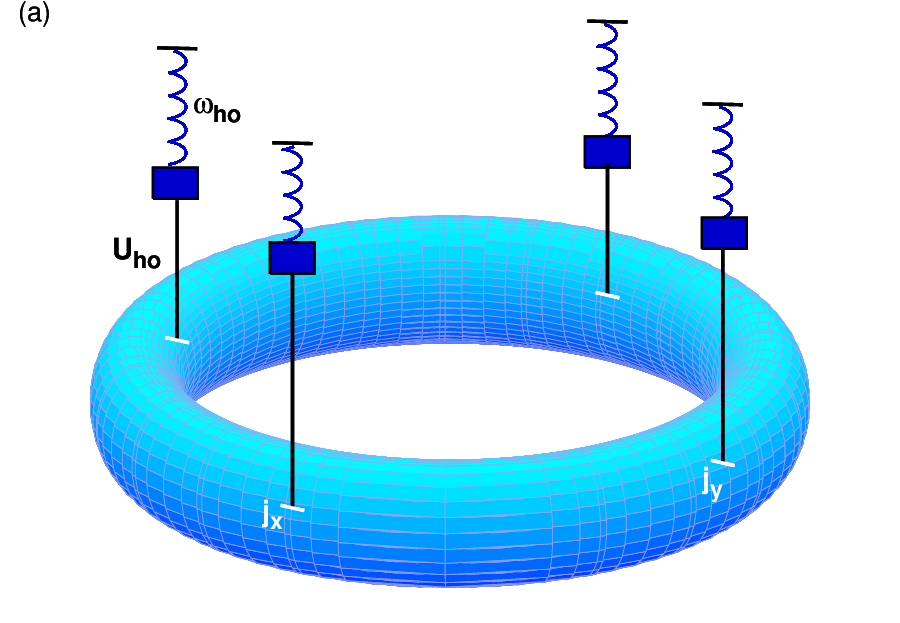}}
\centerline{\includegraphics[scale=0.5]{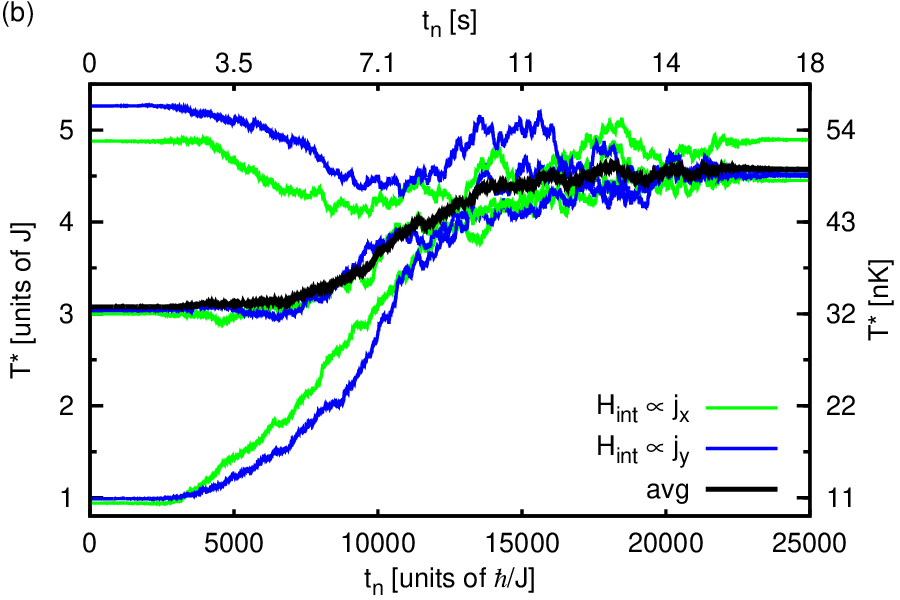}}
\caption{\label{fig:ho_therm} (Color online). (a) Schematic diagram of harmonic oscillator thermometers.  (b) Effective temperature of the oscillators, obtained by evaluating Eqn.~(\ref{effT}) at each time step.  $N = 51407$ atoms. The lattice dimensions are $N_x=126$, $N_y=17$ and  $N_z=5$. In this simulation 512 realizations are used. The turn on and off times are $t_1= 3150\hbar/J$ and $t_2= 21950\hbar/J$; the time scale for both turning on and off the coupling is $\tau_{ho}=1200 \hbar/J$. The initialization temperatures of the oscillators are $T_{0, ho}=1J,3J$ and $5J$. The atomic ensemble is initialized with $T_0/J=1$.  The harmonic trap is ramped on at $t=8050\hbar/J$, with a time constant of $\tau=3200\hbar/J$. After turning off the couplings to the oscillators, the temperature is estimated to be $T^{\ast} = (4.57 \pm 0.09) J = (49.3 \pm  1.0)$ nK.}
\end{figure}

We couple the oscillators to the current 
$
j_{\gv{\alpha}} (\v{r_s}) = - i J l \hbar^{-1}\left( \psi_{\v{r_s}}^\ast \psi_{\v{r_s}+\gv{\alpha}}
																		 - \psi_{\v{r_s}+\gv{\alpha}}^{\ast} \psi_{\v{r_s}} \right) 
$, 
where $\v{r_s}+\gv{\alpha}$ is the nearest-neighbor site either in $x$- or  in $y$- direction. A schematic diagram of the harmonic oscillator thermometers is shown in Fig.~\ref{fig:ho_therm}(a).

We turn the coupling on and off adiabatically slow, using the time dependence 
 $\gamma(t_n)=\left\lbrace\tanh\left[(t_n-t_1)/\tau_{ho}\right] - \tanh\left[(t_n-t_2)/\tau_{ho}\right]\right\rbrace/2$.  The time difference between turn on and turn off times $t_1$ and $t_2$ is chosen long enough to allow the oscillators to equilibrate with the atomic cloud. This was checked by inspecting if  the energy $\expv{E} = \sum_s \expv{H_{0s}}/N_{ho}$   had reached a steady state.

The effective temperature of the cloud is calculated from the expectation value of the energy after decoupling, 
\begin{eqnarray}
T^* & = & \frac{ \hbar \omega_{ho}/2}{\arctanh(\hbar\omega_{ho}/2\expv{E})}.\label{effT}
\end{eqnarray}
The oscillators are initialized according to their Wigner distribution at finite temperature $T_{0, ho}$. This corresponds to sampling from a product of Gaussian distributions with variances $\sigma_x^2 = 1/\left[2\omega_{ho}\tanh(\omega_{ho}/2T_{0,ho}) \right]$ and $\sigma_p^2=\omega_{ho}^2\sigma_x^2$ for $x$ and $p$, respectively. 
As a further check, the initial temperatures of the harmonic oscillators are set to different values. 
We find that the energies of the oscillators converges towards the same steady state value, independent of the initial temperature. This is a further indication that the oscillators have equilibrated with the atomic ensembles, and that the measured temperature is a good estimate of the atomic ensemble temperature.  

In the numerical results reported here, six harmonic oscillators, equally spaced along the ring, are coupled to the atomic current at the trap center in the x- and y-directions,  at the location of maximum density.  Half of the oscillators are coupled to the x-current and the other half are coupled to the y-current.  In Fig.~\ref{fig:ho_therm}(b) an example for the time evolution of the effective temperature of the six oscillators is plotted.
For this example and throughout the paper, the harmonic oscillator parameters are $U_{ho}=0.008J$, $\omega_{ho}=2J$ and $\tau_{ho}= 1200 \hbar/J$.  These parameters were chosen in a way to minimize both the computational time and the coupling strength. We check that the oscillators do not introduce measurable heating of the atoms.  
 The green (blue) lines correspond to the oscillators that are coupled to the x- (y-) current.   Each line represents an average of 512 realizations.  The thick black line is an average over all of the oscillators and all of the realizations.  As can be seen from the plot, the oscillators initialized to different temperature converge to a single temperature, the effective temperature stabilizes in time and the final temperature of the oscillators coupled to the x-current is indistinguishable from those coupled to the y-current.   The oscillators initialized to $T_{0,ho}=5$J cool initially and then heat up again as the trap is ramped on.  The final temperature for the data in Fig.~\ref{fig:ho_therm}(b) is $T^\ast=(4.57 \pm 0.09) J = (49.3 \pm  1.0)$ nK.

We have checked that the final temperature of the atom cloud does not depend on the parameters chosen, within the error, indicating that any heating associated with the oscillators is negligible.  We consider this method presented here to be generally applicable to a wide range of TWA simulations. 

\section{Estimating the Chemical Potential}\label{appendix:chem_potential}
%
 In a homogeneous system in equilibrium and with a well-defined dimension, the chemical potential is a well-defined quantity. However, in our simulations we consider a trapped system out of equilibrium and with regions of varying dimensionality, such as the bulk and the barrier region. Despite this, the energy scale of a `local' chemical potential is a useful quantity in discussing the behavior of the system, even though it only has an approximate meaning. 

Because of the different types of data available from the simulations and the experiment, we employ several different methods of estimating the global and local chemical potential.  These approximations are:  (1) $\mu(x)$, which estimates the chemical potential based on the density at  location $(x,0,0)$, within the Thomas-Fermi approximation; (2) $\mu_0$, the average-density-based approximation, which is based on the density at the trap minimum averaged around the ring; (3) $\mu_N$, the atom-number-based approximation, which is calculated from the total number of atoms, assuming a Thomas-Fermi distribution in both $y$- and $z$-direction; (4)  $\mu_{2D}(x)$, a column-density-based approximation, in the limit that locally the system is two-dimensional (2D), calculated from the column density at $(x,0)$ and (5) $\mu_{1D}(x)$, based on the density integrated over both $y$ and $z$, which is applicable when the system is one-dimensional. The approximations $\mu_0$ and $\mu_N$ estimate the global chemical potential, while the others estimate the `local' chemical potential.  The approximations $\mu_N$ and $\mu_{2D}$ are specifically calculated to compare the numerical and experimental results.  

We briefly outline the method for determining each quantity.  The energy functional of the GPE is \cite{gross_structure_1961, *pitaevskii_vortex_1961}:
\begin{align}
E[\psi] = \int d\v{r} \Bigg[ &-\frac{\hbar^2}{2m}|\grad\psi(\v{r})|^2 + \frac{g}{2} | \psi(\v{r}) |^4  \notag\\  
										 & + V(\v{r})|\psi(\v{r})|^2 \Bigg], \label{eqn:gpe_functional}
\end{align}
where $V(\v{r})$ is the trapping potential, $V(\v{r}) = \frac{1}{2}m \left( \omega_y^2 y^2 + \omega_z^2 z^2 \right)$.
To determine the equilibrium state the GPE energy functional is minimized, which gives 
\begin{equation*}
\left[- \hbar^2/(2m)|\grad\psi(\v{r})|^2 + g|\psi(\v{r})|^2 + V(\v{r}) \right] \psi(\v{r}) = \mu \psi(\v{r}),
\end{equation*}
 where  $\mu$ is introduced as a Lagrange multiplier. Within the Thomas-Fermi approximation, the kinetic energy term is neglected and the density is
\begin{equation}\label{eqn:n(mu)}
n(\v{r}) = |\psi(\v{r})|^2 = g^{-1} \left( \mu - V(\v{r}) \right).
\end{equation}
 At the trap minimum, the local chemical potential is 
$$
\mu({x})= g n(x,0,0). 
$$
This approximation applies when the chemical potential is greater than the trapping energies, $\mu > \hbar \omega_y, \hbar \omega_z$.

To improve this estimate in the numerical evaluation, we calculate the average density at the trap minimum, where $V(\v{r})=0$,  
$$
\mu_{0}\equiv N_x^{-1} \sum_x \mu(x).
$$
We calculate this quantity in the absence of the barrier,  to estimate the bulk chemical potential. 

The chemical potential can also be determined by summing the density over all space in Eqn.~(\ref{eqn:n(mu)}), using the condition
$\int d{\v{r}}n(\v{r}) = N$, and solving for the chemical potential as a function of the total number of atoms.
The total-number-based approximation for the chemical potential is
\begin{equation*}
\mu_N = \left(\frac{g N m \omega_y\omega_z}{\pi L_x}\right)^{1/2}
		 = \left(\frac{Ul N \hbar^2\omega_y\omega_z}{2\pi L_x J}\right)^{1/2}.
\end{equation*}
This  expression assumes a Thomas-Fermi profile in $y$- and $z$-direction.  This quantity was used in the experiment in Ref. \cite{ramanathan_superflow_2011}, and is used here to compare the simulations and the experimental results in Fig.~\ref{fig:muc_Vb}.

Furthermore, the chemical potential at the barrier was calculated from the two-dimensional column density at the barrier in Ref. \cite{ramanathan_superflow_2011}.  At the barrier, the density is sufficiently suppressed that the local chemical potential is less than the harmonic confinement in $z$-direction, so that it is effectively 2D. The wavefunction is separable, $\psi(\v{r}) = \psi_{TF}(x,y)\psi_{ho}(z)$, into a product of the harmonic oscillator ground state 
$\psi_{ho}(z) = \left(1/\pi l_z^2 \right)^{1/4}\exp\left(-z^2/2l_z^2\right) $, where $l_z=\sqrt{\hbar/m\omega_z}$, and  the Thomas-Fermi distribution in $x$- and $y$-direction. The harmonic oscillator ground state is normalized as $\int d{z} |\psi_{ho}(z)|^2 =1$ so that $|\psi_{TF}(x,y)|^2$ corresponds to the column density measured in the experiment. We substitute $\psi_{TF}(x,y), \psi_{ho}(z)$ into Eqn.~(\ref{eqn:gpe_functional}) and integrate over $z$ to get 
\begin{align*}
E[\psi] = \int dx dy  \bigg[& -\frac{\hbar^2}{2m} |\grad \psi_{TF}|^2  + \frac{1}{2} \left(m\omega_y^2 y^2 + \hbar\omega_z \right) |\psi_{TF}|^2  \\
										 & + \frac{g}{2} \frac{1}{\sqrt{2\pi} l_z} |\psi_{TF}|^4\bigg].
\end{align*}
Again, we ignore the kinetic energy term, and subtract the constant offset to the chemical potential due to the harmonic oscillator energy, $\hbar\omega_z/2$. We minimize the total energy, while $\mu$ controls the total number of particles.  
The resulting  2D column density is 
\begin{align*}
n_{2D}(x,y) =|\psi_{TF}|^2  = \frac{1}{g_{2D}} \left[\mu - \frac{1}{2}m\omega_y^2 y^2 \right]
\end{align*}
where $g_{2D} = g/\sqrt{2\pi}l_z$.  Solving this expression for the chemical potential in terms of the peak column density, $n_{2D}$, which occurs at $y=0$, yields
\begin{equation*}
\mu_{2D}(x) = g_{2D}n_{2D}(x,0).   
\end{equation*}
 The chemical potential for the quasi-1D case can be determined in a similar manner, by replacing the full wavefunction with $\psi(\v{r}) = \psi(x)\psi_{ho, y}(y)\psi_{ho, z}(z)$.  The resulting chemical potential is 
 $$
 \mu_{1D}(x) = g_{1D}n_{1D}(x)
 $$
where $g_{1D} = g/(2\pi l_{z} l_{y})$, and $n_{1D}$ is the density integrated along $y$- and $z$-direction, $n_{1D}(x) = \sum_{y,z} n(x,y,z)$.
 
\section{Speed of Sound}\label{appendix:sound_speed}
%
An important dynamic quantity of a condensed Bose gas is the phonon velocity. For a homogeneous, weakly interacting Bose gas in 3D it is given by $c_s = \sqrt{gn/m}$. However, the system that we consider here has a spatially varying density, even to the degree that the dimension of the system varies, for example in the vicinity of the barrier. We therefore introduce several limiting expressions for the phonon velocity similar to the previous section, in which several limits for the chemical potential were discussed.

For a Bose condensate in a cylindrical geometry, the phonon velocity is approximately given by
$$
c_{3D}(x) = \sqrt{\frac{gn(x,0,0)}{2m}} = \sqrt{\frac{\mu(x)}{J}}\frac{Jl}{\hbar}.
$$
The density $n$ has been replaced by the average density over the cylinder, $\bar{n} = n(x,0,0)/2$, where we assume a Thomas-Fermi profile in $y$- and $z$-direction, see e.g. \cite{zaremba_sound_1998, mathey_phase_2010}. This is valid when the chemical potential is greater than the harmonic confinement energy in the transverse directions. When $\hbar\omega_y <\mu(x) < \hbar\omega_z$, the system is quasi-2D and the local speed of sound is
$$
c_{2D}(x) = \sqrt{\frac{2g_{2D} n_{2D}(x,0)}{3m}} 
				= \sqrt{\frac{4 \mu_{2D}(x)}{3 J} } \frac{Jl}{\hbar}.	
$$
This can be derived from the low-energy excitation spectrum within the Bogoliubov de Genes approximation, as in Refs.~\cite{ohberg_low-energy_1997,mathey_phase_2010},  starting from the 2D Hamiltonian. It was obtained in Ref.~\cite{stringari_dynamics_1998} within a hydrodynamic approach. The key step is to average the column density, $n_{2D}(x,y) = (\mu_{2D} - \frac{1}{2}m\omega_y^2 y^2)/g_{2D}$ over $y$:
$ \bar{n}_{2D} = \int dy n_{2D}(x,y) = 2 n_{2D}(x,0)/3$.

At the critical barrier, the condition $\hbar\omega_y < \mu(x) < \hbar\omega_z$ is fulfilled for all of the simulations using the truncated Wigner approximation reported in this paper.  Thus we use $c_{2D}$ as the best approximation at the barrier.  Additionally, the experimental data is rescaled and the critical velocity is compared with $c_{2D}$ instead of $c_{3D}$, as was originally done in \cite{ramanathan_superflow_2011}.  

When $\mu(x) <\hbar\omega_y, \hbar\omega_z$, the system is quasi-1D and the local speed of sound is the same as in the homogenous case, except with the 1D interaction parameter and 1D density, 
$$
c_{1D}(x) 	= \sqrt{\frac{g_{1D}n_{1D}(x) }{m}} 
			 	= \sqrt{\frac{2\mu_{1D}(x) }{J}} \frac{Jl}{\hbar}.
$$
The system is quasi-1D at the critical barrier  for the GPE simulations. We then compare the local velocity to $c_{1D}$ for the GPE data Figs.~\ref{fig:jT_TWA_GPE} and \ref{fig:vc_cs}(a). 

We note that the cross-over from the quasi-1D to the quasi-2D regime is not a sharp transition. We find that in the intermediate regime of $\mu(x) \approx \hbar \omega_y$, $c_{1D}$ is typically $\sim10\text{-}15\%$ lower than $c_{2D}$.

\bibliography{sfdecay_twa}

\end{document}